\documentclass[aps,prd, twocolumn]{revtex4-1}
\usepackage{amsmath, amssymb, graphicx, color}
\def\aap{Astron. Astrophys.}

\def\mnras{MNRAS}
\def\be{\begin{equation}}
\def\ee{\end{equation}}
\def\figwidth{0.47\textwidth}

\begin{document}
\title{Statistics of Thawing K-essence Dark Energy Models}

\author{Zhiqi Huang}
\affiliation{School of Physics and Astronomy, Sun Yat-sen University, 2 Daxue Road, Zhuhai, CHINA}
\email{huangzhq25@mail.sysu.edu.cn}

\date{\today}

\begin{abstract}
  
 K-essence is a minimally-coupled scalar field whose Lagrangian density $\mathcal{L}$ is a function of the field value $\phi$ and the kinetic energy $X=\frac{1}{2}\partial_\mu\phi\partial^\mu\phi$. In the thawing scenario, the scalar field is frozen by the large Hubble friction in the early universe, and therefore initial conditions are specified. We construct thawing k-essence models by generating Taylor expansion coefficients of $\mathcal{L}(\phi, X)$ from random matrices. From the ensemble of randomly generated thawing k-essence models, we select dark energy candidates by assuming negative pressure and non-growth of sub-horizon inhomogeneities. For each candidate model the dark energy equation of state function is fit to the Chevallier-Polarski-Linder parameterization $w(a) \approx w_0+w_a(1-a)$, where $a$ is the scale factor. The thawing k-essence dark models distribute very non-uniformly in the $(w_0, w_a)$ space. About 90\% models cluster in a narrow band in the proximity of a slow-roll line $w_a\approx  -1.42 \left(\frac{\Omega_m}{0.3}\right)^{0.64}(1+w_0)$, where $\Omega_m$ is the present matter density fraction. This work is a proof of concept that  for a certain class of models very non-uniform theoretical prior on $(w_0, w_a)$ can be obtained to improve the statistics of model selection.
 
\end{abstract}

\maketitle

\section{Introduction \label{sec:intro}}

The cosmological concordance Lambda cold dark matter ($\Lambda$CDM) model, where $\Lambda$ stands for the cosmological-constant interpretation of dark energy, provides a remarkably good fit to most of the current cosmological data sets~\cite{Planck18Params,BOSS2021,DES3yr,Pantheon}. Although the recent debates on Hubble tension~\cite{Riess21} raise some doubts,  $\Lambda$CDM is at least confirmed to be a good  phenomenological approximation to reality.

If the constant $\Lambda$ term in Hilbert-Einstein action is indeed a phenomenological approximation, one may wonder whether a similar but dynamic scalar degree of freedom could be a more accurate description of reality. Because high-derivative theories typically suffer from Ostrogradsky instability~\cite{Ostrogradsky}, it is often assumed that the scalar-field Lagrangian density is a function of the field value $\phi$ and the kinetic energy $X = \frac{1}{2}\partial_\mu\phi\partial^\mu\phi$. Such minimal scalar-field models, or in modern terminology, k-essence models were initially proposed to resolve the coincidence problem of dark energy, which questions why the dark energy density is close to the matter density at present epoch~\cite{Kessence_APC}. The proposed models contain carefully tuned parameters that lead to attractor solutions, where dark energy evolves dynamically since early universe and settles down to a density that is close to the matter density today. It was later understood that k-essence is not a very successful solution of coincidence problem, as it requires additional fine-tuning, and the basins of attraction represent only a small region of the phase space~\cite{Coincidence}. The philosophy here now is to take k-essence as a phenomenological extension of $\Lambda$. Now that attractor solutions are no longer our pursuit, the most natural assumption might be the absence of early-universe dynamics. The so-called thawing scenario, where the scalar field is frozen by the large Hubble friction in the early universe, will be implicitly assumed throughout this work.

Many future cosmological surveys aim to reconstruct the Chevallier-Polarski-Linder (CPL) parameterization of the dark energy equation of state (EOS)~\citep{Chevallier:2000qy,Linder:2002et}
\be
w(a) = w_0+w_a(1-a), \label{eq:w0wa}
\ee
where $a$ is the scale factor normalized to unity today. The CPL parameterization can be regarded as a truncated Taylor expansion at redshift zero, which may not work well at high redshift. However, it is a useful approximation, because in the standard scenario dark energy is negligible at high redshift. In this view, the intercept parameter $w_0$ approximates the present value of dark energy EOS, and the slope parameter $w_a$ represents the low-redshift running. Many dark energy models are then projected, in most cases approximately, onto the $w_0$-$w_a$ chart before they are confronted with observational data. The cosmological constant maps to $(w_0, w_a) = (-1, 0)$, for instance. In most applications, independent and uniform priors are applied to $w_0$ and $w_a$. For a specific class of models, however, $w_0$ and $w_a$ may be correlated. For instance, when a canonical scalar field (quintessence) slowly rolls down from a flat potential, its current kinetic energy ($\sim 1+w_0$) is correlated with its recent rolling history ($\sim w_a$). Indeed, approximate functional forms of $w(a)$ for slow-roll quintessence have been obtained in Refs.~\cite{Crittenden2007, Scherrer2008, Chiba2009, WZ_Huang10, WZ_Miao18}. For non-slowroll quintessence or more general k-essence, the complexity arises and only very few simple cases were studied~\cite{WZ_Chiba, WZ_Kehayias}. Because  for k-essence it becomes impractical to find a definitive functional form of $w(a)$, we take a step back and look for a statistical description. Our work is similar in approach to Refs.~\cite{HP2007,Marsh14,Garcia19}, which studied the statistics of quintessence models rather than k-essence.

Throughout the paper we work with natural units $c=\hbar=1$ and a spatially flat Friedmann-Robertson-Walker (FRW) background metric with scale factor $a(t)$, where $t$ is the cosmological time. Derivative with respect to $t$ is denoted as an overhead dot. The Hubble parameter is defined as $H \equiv \frac{\dot a}{a}$, whose current value, the Hubble constant, is denoted as $H_0$ or $100  h\;\mathrm{km\, s^{-1}\mathrm{Mpc}^{-1}}$. We define the reduced Planck mass $M_p \equiv \frac{1}{\sqrt{8\pi G_N}}$, where $G_N$ is Newton's Gravitational constant. The dimensionless density parameters $\Omega_m$, $\Omega_{\gamma}$, $\Omega_{\nu i}$ ($i=1,2, 3$), and $\Omega_\phi$ are defined as the present fractional background density of matter, radiation, $i$-th neutrino, and the thawing k-essence dark energy, respectively. The critical energy density is given by $\rho_c \equiv 3H_0^2M_p^2$.

\section{Statistics of CPL fittings}

We start with a general two-variable function $\mathcal{L} = p(\phi, X)$. In action
\be
S = \int d^4x \sqrt{-g}\, p(\phi, X),
\ee
varying the field value $\phi$ yields its equation of motion (EOM)
\be
\frac{1}{\sqrt{-g}}\partial_\mu\left(\sqrt{-g}\frac{\partial p}{\partial X}\partial^\mu\phi\right) - \frac{\partial p}{\partial \phi} = 0,
\ee
where $g$ is the determinant of the spacetime metric $g_{\mu\nu}$.

The EOM is equivalent to
\be
    \frac{\partial p}{\partial X}\square \phi + \left(\frac{\partial^2 p}{\partial X^2}\partial_\mu\partial^\nu\phi\partial_\nu\phi + \frac{\partial^2p}{\partial X\partial \phi}\partial_\mu\phi\right)\partial^\mu\phi - \frac{\partial p}{\partial \phi} = 0.
    \ee
where $\square$ is the four-dimensional covariant Laplacian operator.

In the cosmological context, the EOM can be split into a background component
\be
\ddot\phi + 3Hc_s^2\dot\phi + S(\phi, X) = 0, \label{eq:bg}
\ee
and a perturbation component that describes how the inhomogeneities of $\phi$ evolve. 
The source term in Eq.~\eqref{eq:bg} is given by
\be
S(\phi, X)\equiv \frac{\frac{\partial \rho}{\partial\phi}}{\frac{\partial \rho}{\partial X}}, \label{eq:S}
\ee
where
\be
\rho(\phi, X) = 2X\frac{\partial p}{\partial X}-p \label{eq:rho}
\ee
is the energy density of k-essence. The effective sound speed squared is defined as
\be
c_s^2 = \frac{\frac{\partial p}{\partial X}}{\frac{\partial \rho}{\partial X}}. \label{eq:cs2}
\ee
We do not consider models with $c_s^2<0$, which implies ultraviolet instability.

On sub-horizon scales,  we may approximately use the linear perturbation equation in a perfect FRW background,
\be
\ddot{\delta\phi} + \left(3Hc_s^2 +  \frac{\partial S}{\partial X}\dot\phi\right)\dot{\delta\phi} + \left(\frac{\partial S}{\partial\phi} - \frac{c_s^2}{a^2}\nabla^2\right)\delta\phi  = 0, \label{eq:pert}
\ee
to qualitatively describe the clustering behavior of dark energy. In practice, we evolve Eq.~\eqref{eq:pert} for a pivot scale $k_{\rm pivot}=0.05 h\, \mathrm{Mpc}^{-1}$ and discard the models where the growth factor of dark energy density perturbations ($\lvert\frac{\delta\rho}{\rho}\rvert$, initially normalized to unity) exceeds $100$. The pivot scale and the cutoff of density growth do not correspond to a particular observational constraint. Our philosophy here is to exclude models with large sub-horizon inhomogeneities, which typically cannot be approximately treated as a $w_0$-$w_a$ model and hence is beyond the scope of the present work.

We now assume that $p(\phi, X)$ can be expanded in Taylor series of $X$,
\be
p(\phi, X) = \rho_c\sum_{i=0}^\infty \frac{p_i(\phi)}{i!}\left(\frac{X}{\rho_c}\right)^i. \label{eq:p_expand}
\ee
It follows that the energy density is
\be
\rho(\phi, X) = \rho_c\sum_{i=0}^\infty  \frac{(2i-1)p_i(\phi)}{i!}  \left(\frac{X}{\rho_c}\right)^i. \label{eq:rho_expand}
\ee

The k-essence action is invariant under a change of variable, in particular, a translation $\phi\rightarrow \phi + const.$. Without loss of generality, we take the early-time field value as the zero point and set the thawing initial conditions 
\be
\left.\phi\right\vert_{t\rightarrow 0} = \left.\dot\phi\right\vert_{t\rightarrow 0} = 0. \label{eq:ini}
\ee
The initial conditions~\eqref{eq:ini}  does not guarantee the thawing scenario. We still need to discard the cases where the early-universe Hubble friction is not large enough to freeze the field. In practice, we discard the models with $1+w=1+p(\phi, X)/\rho(\phi,X)>0.01$ at any redshift beyond $z_{\rm CMB}= 1089$.

We now expand $p_i(\phi)$ ($i=0,1,2,\ldots$), which is dimensionless, as
\be
p_i(\phi) =  \sum_{j=0}^\infty \frac{V_{ij}}{j!}\left(\frac{\phi}{\sqrt{3}M_p}\right)^j.\label{eq:V_expand}
\ee
A model can then be constructed by randomly generating the matrix $V_{ij}$. In practice, we work with a truncated finite-size matrix $V_{ij}$ ($0\le i, j < n$), where the integer $n \gg 1$ is a truncation order. 

In the FRW universe, the Hubble parameter is given by
\be
\frac{H}{H_0} = \sqrt{\Omega_ma^{-3} + \left[\Omega_\gamma + \sum_{i=1}^3 \Omega_{\nu,i}\frac{\mathrm{I}_{\rho}\left(\frac{m_{\nu,i} a}{T_{\rm CNB}}\right)}{\mathrm{I}_{\rho}\left(\frac{m_{\nu,i}}{T_{\rm CNB}}\right)}\right]a^{-4} + \frac{\rho(\phi, X)}{\rho_c}}, \label{eq:H}
\ee
where $m_{\nu, i}$ is the neutrino mass of the $i$-th specie, and $T_{\rm CNB}=1.95\mathrm{K}$ is the effective temperature for neutrino momentum distribution. The neutrino density integral is
\begin{equation}
  \mathrm{I}_\rho(\lambda) \equiv  \frac{1}{2\pi^2}\int_0^\infty \frac{x^2\sqrt{x^2+\lambda^2}}{e^x+1}dx.
\end{equation}
Three neutrino species with masses $0.05\mathrm{eV}$, $0.009\mathrm{eV}$, $0.001 \mathrm{eV}$ are assumed by default.

We use the second Friedmann equation
\begin{eqnarray}
\frac{\ddot a}{a} &=& -\frac{H_0^2}{2}\left[\Omega_ma^{-3}+ \frac{\rho(\phi, X)+3p(\phi, X)}{\rho_c}  \right. \\
&+& \left. \left( 2\Omega_\gamma  +  \sum_{i=1}^3 \Omega_{\nu,i}\frac{\mathrm{I}_{\rho}\left(\frac{m_{\nu,i} a}{T_{\rm CNB}}\right)+ 3\mathrm{I}_p\left(\frac{m_{\nu,i} a}{T_{\rm CNB}}\right)}{\mathrm{I}_{\rho}\left(\frac{m_{\nu,i}}{T_{\rm CNB}}\right)}\right)a^{-4} \right], \nonumber
\end{eqnarray}
to evolve the scale factor,  where the neutrino pressure integral is given by
\begin{equation}
  \mathrm{I}_p(\lambda) \equiv  \frac{1}{6\pi^2}\int_0^\infty \frac{x^4}{\sqrt{x^2+\lambda^2}\left(e^x+1\right)}dx.
\end{equation}
The first Friedmann equation (i.e., Eq.~\eqref{eq:H}) is an energy conservation equation that can be used to check the numeric accuracy. We tune the numeric step size such that the relative error of Eq.~\eqref{eq:H} is controlled below $10^{-4}$. 

In the numeric code we evolve $\frac{H}{H_0}$ as a function of redshift. Energy conservation does not guarantee that both sides of Eq.~\eqref{eq:H} are equal to unity, as they should be, when we stop the evolution at redshift zero. In other words, the model parameters $V_{ij}$ ($0\le i, j < n$) contain a redundant degree of freedom that need to be tuned to guarantee self-consistency. It follows from Eqs.~(\ref{eq:rho_expand}-\ref{eq:V_expand}) that the early dark energy density $\left.\rho\right\vert_{t\rightarrow 0} = - V_{00}\rho_c$. For dark energy model without early dynamics, the early dark energy density $\left.\rho\right\vert_{t\rightarrow 0}$ should not be too many orders of magnitude away from the current dark energy density $\sim O(1) \rho_c$. Thus, we choose $V_{00}$ as the parameter to be tuned. In practice, we numerically solve $V_{00}$ with a logarithmic binary search method in a range $-10^2<V_{\rm 00}<-10^{-2}$.

In summary, our numeric scheme to generate dark energy candidates is as follows.
\begin{enumerate}
\item{Randomly generate $V_{ij}$ ($0\le i, j < n$, but excluding $V_{00}$) from independent Gaussian distributions $P\left(V_{ij}\right)\propto e^{-V_{ij}^2/(2\sigma^2)}$, where $\sigma\gtrsim O(1)$ is a fixed sampling width.}
\item{Using initial conditions~\eqref{eq:ini}, EOM~\eqref{eq:bg} and self-consistency requirement ($\frac{H}{H_0}\rvert_{z=0}=1$), numerically solve $V_{00}$. If no solution exits, discard the model and go back to step 1.}
\item{From the background solution compute $w(a)$ and $c_s^2(a)$. If $w> -\frac{1}{3}$ or $c_s^2<0$ at any time, or $\lvert 1+w\rvert > 0.01$ at any early time (redshift $z> z_{\rm CMB} = 1089$), discard the model and go back to step 1.}
\item{Evolve the perturbation Eq.~\eqref{eq:pert}. If the growth factor of dark energy density perturbations at pivot scale exceeds $100$ at any redshift, discard the model and go back to step 1.}
\item{Fit $w(a) = w_0+w_a(1-a)$ ($a\in [a_{\rm late}, 1]$) with a least-square method.}
\end{enumerate}
A sample Fortran code is shared at \url{http://zhiqihuang.top/codes/scan_kessence.tar.gz} to allow reproduction of our results.

\begin{figure}
  \includegraphics[width=\figwidth]{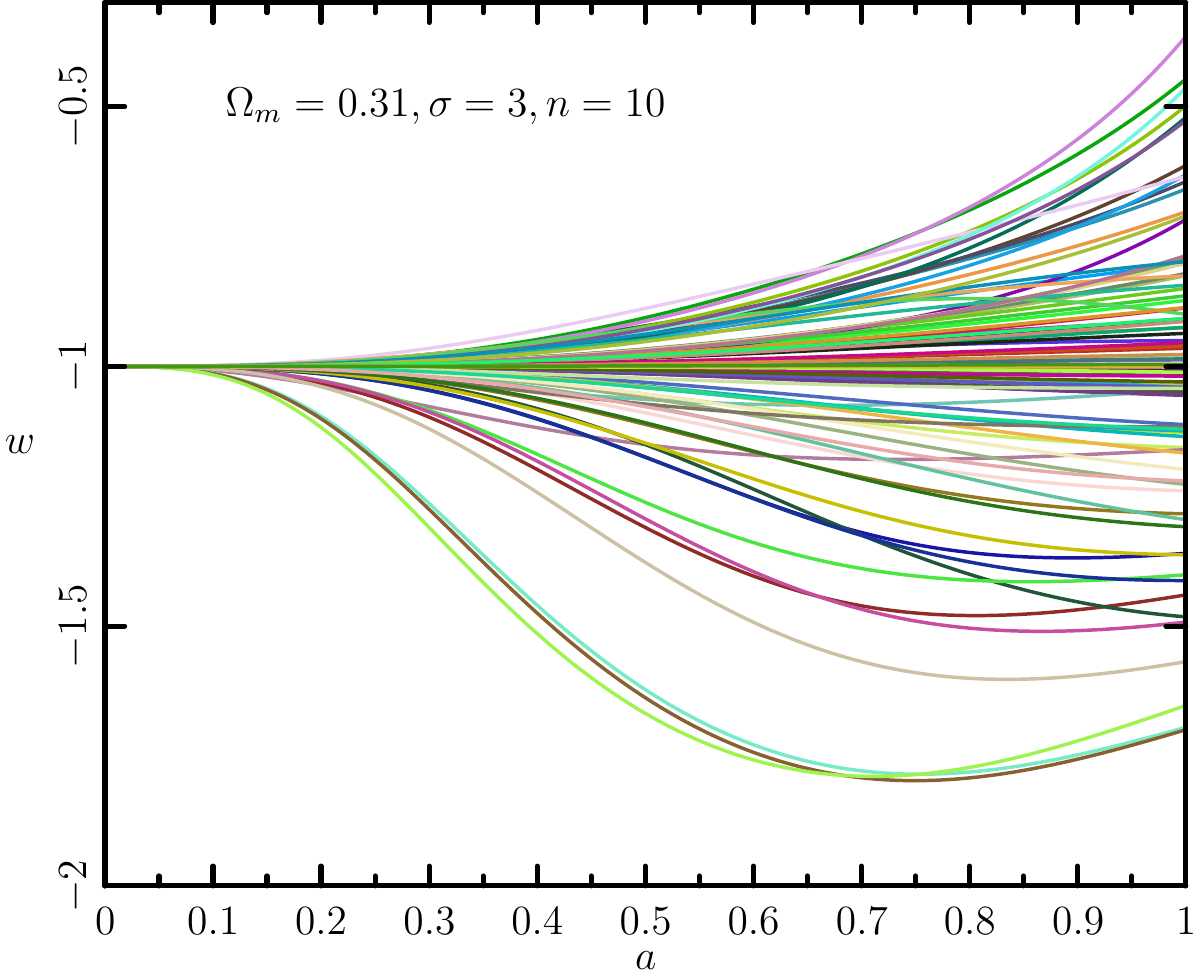}

  \includegraphics[width=\figwidth]{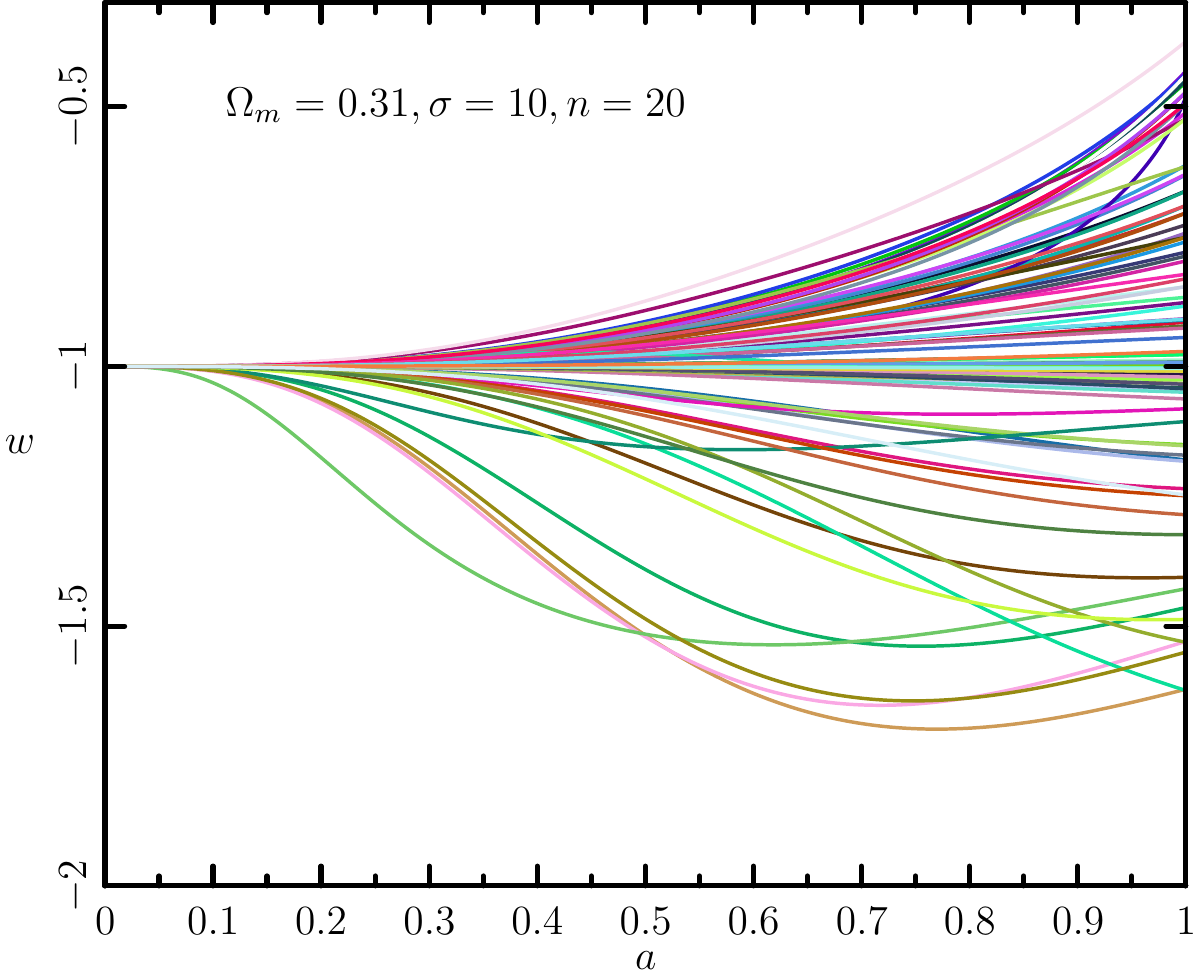}
  \caption{EOS of randomly sampled thawing k-essence dark energy candidates. Each panel contains 100 random samples. \label{fig:trajs}}
\end{figure}
\begin{figure*}
  \includegraphics[width=0.31\textwidth]{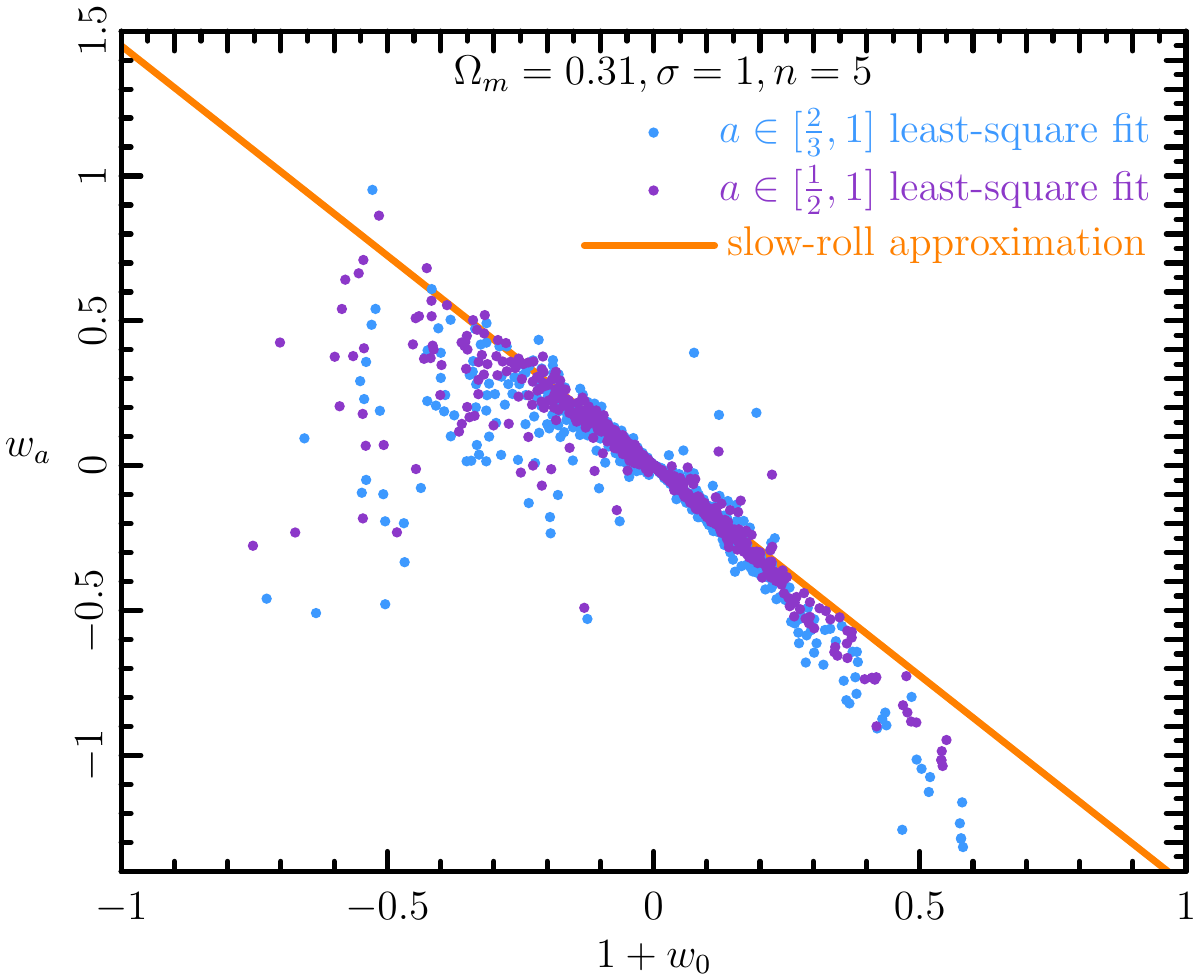}  \includegraphics[width=0.31\textwidth]{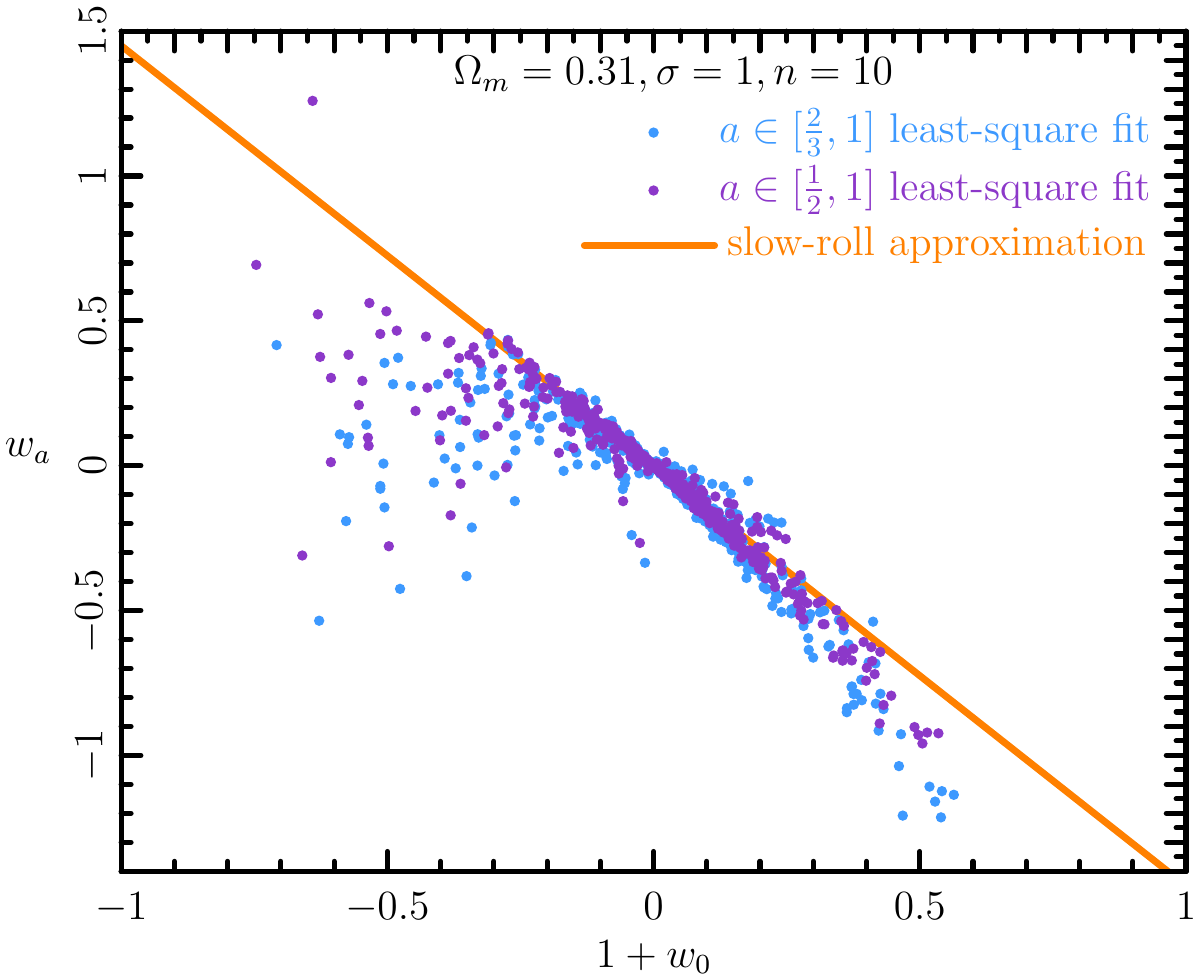}  \includegraphics[width=0.31\textwidth]{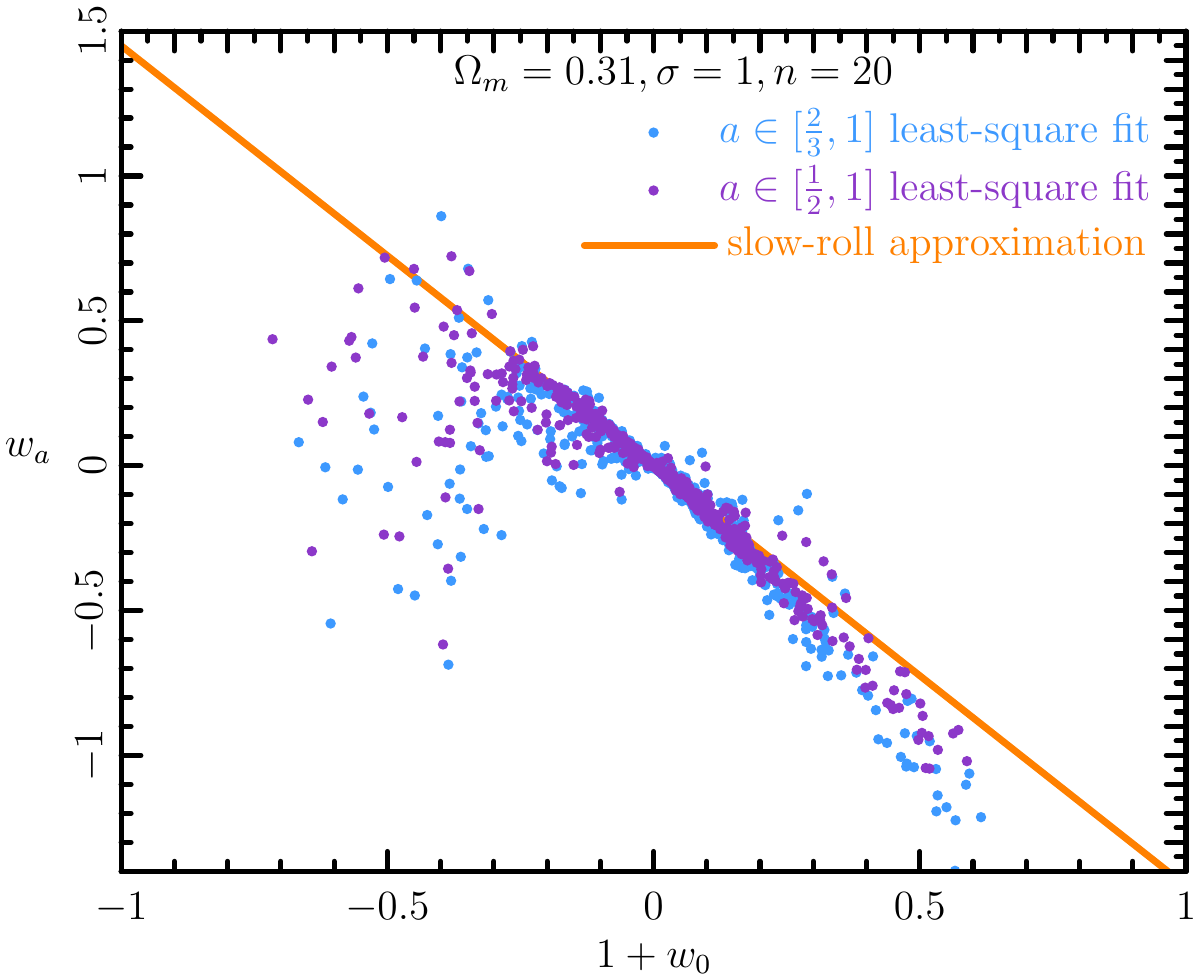} \\
  \includegraphics[width=0.31\textwidth]{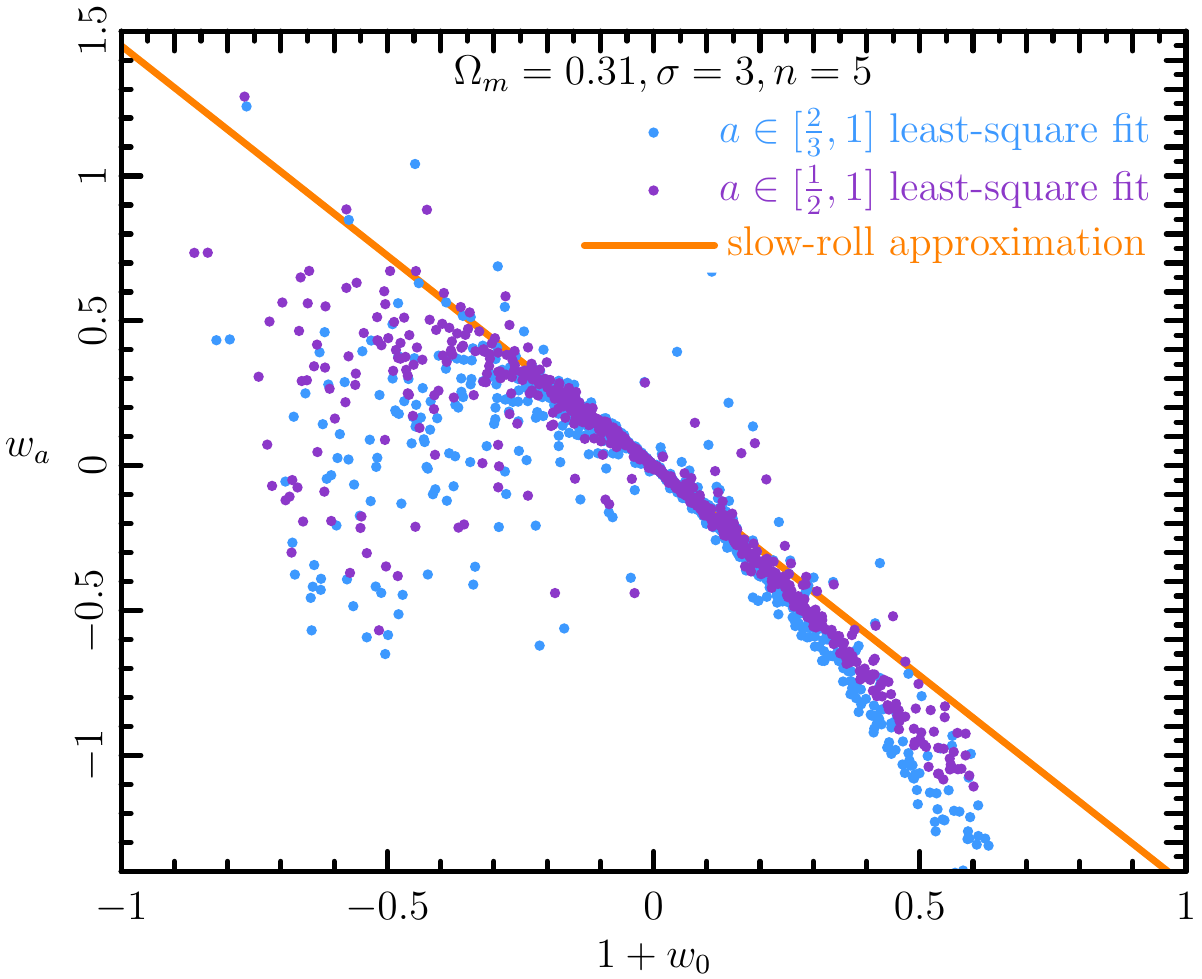}  \includegraphics[width=0.31\textwidth]{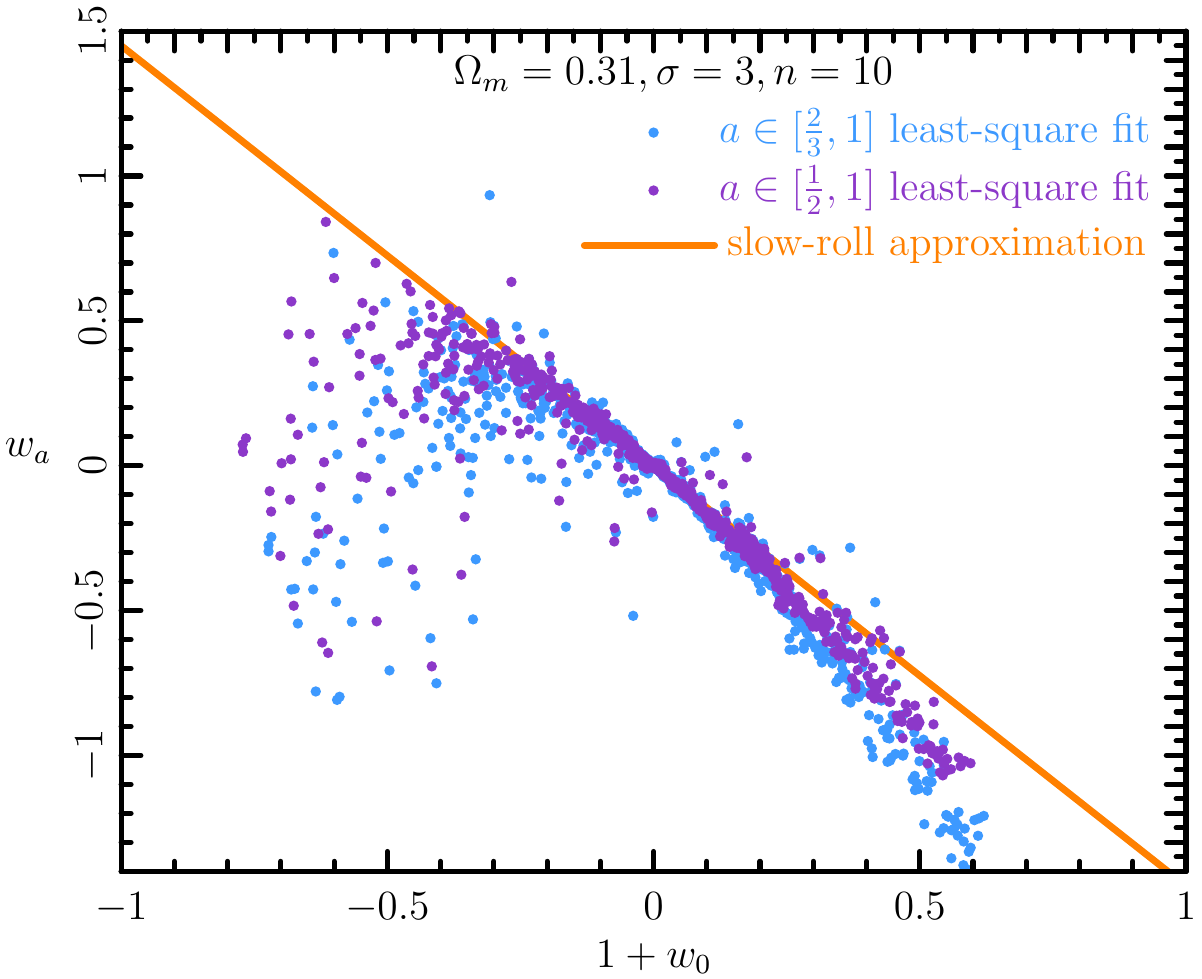}  \includegraphics[width=0.31\textwidth]{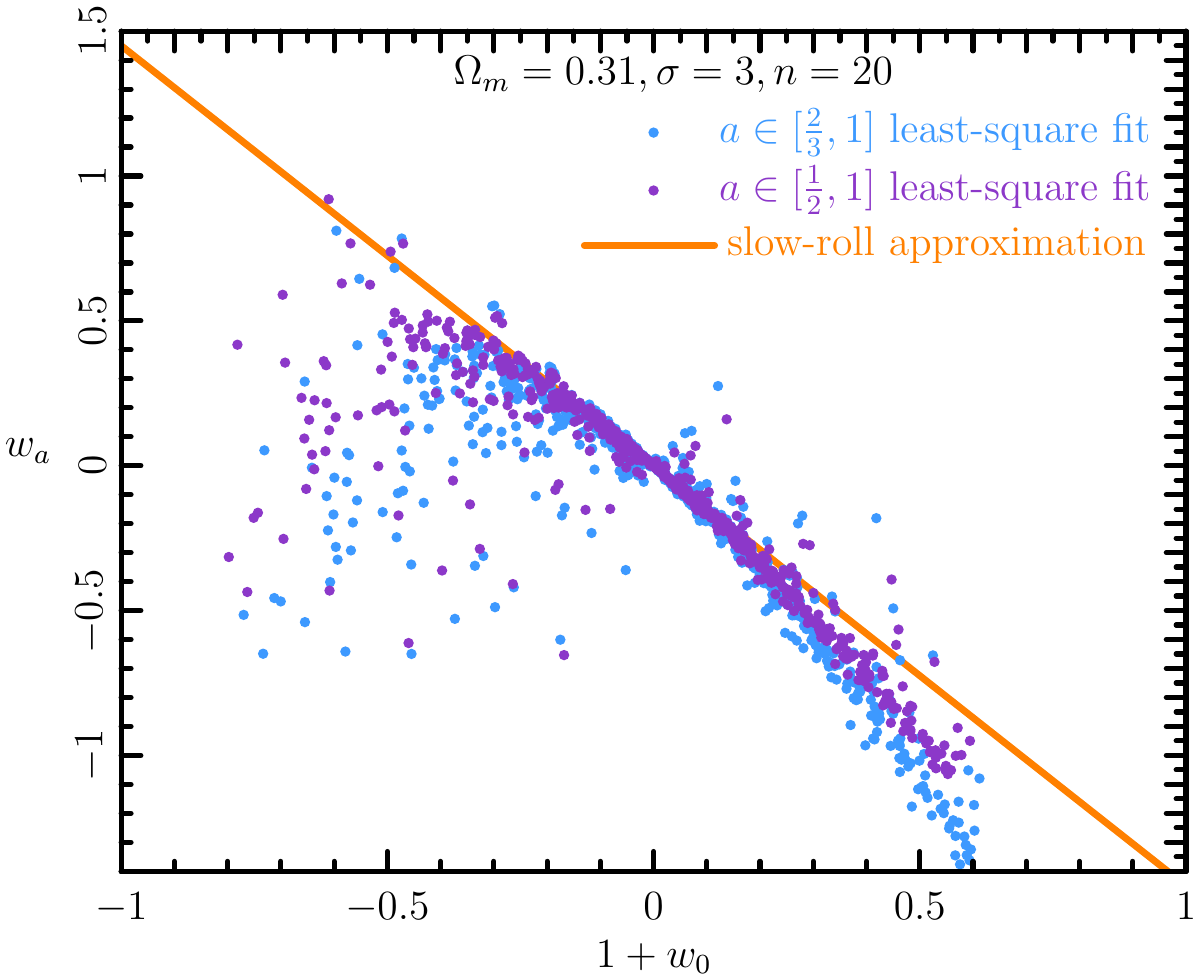} \\
  \includegraphics[width=0.31\textwidth]{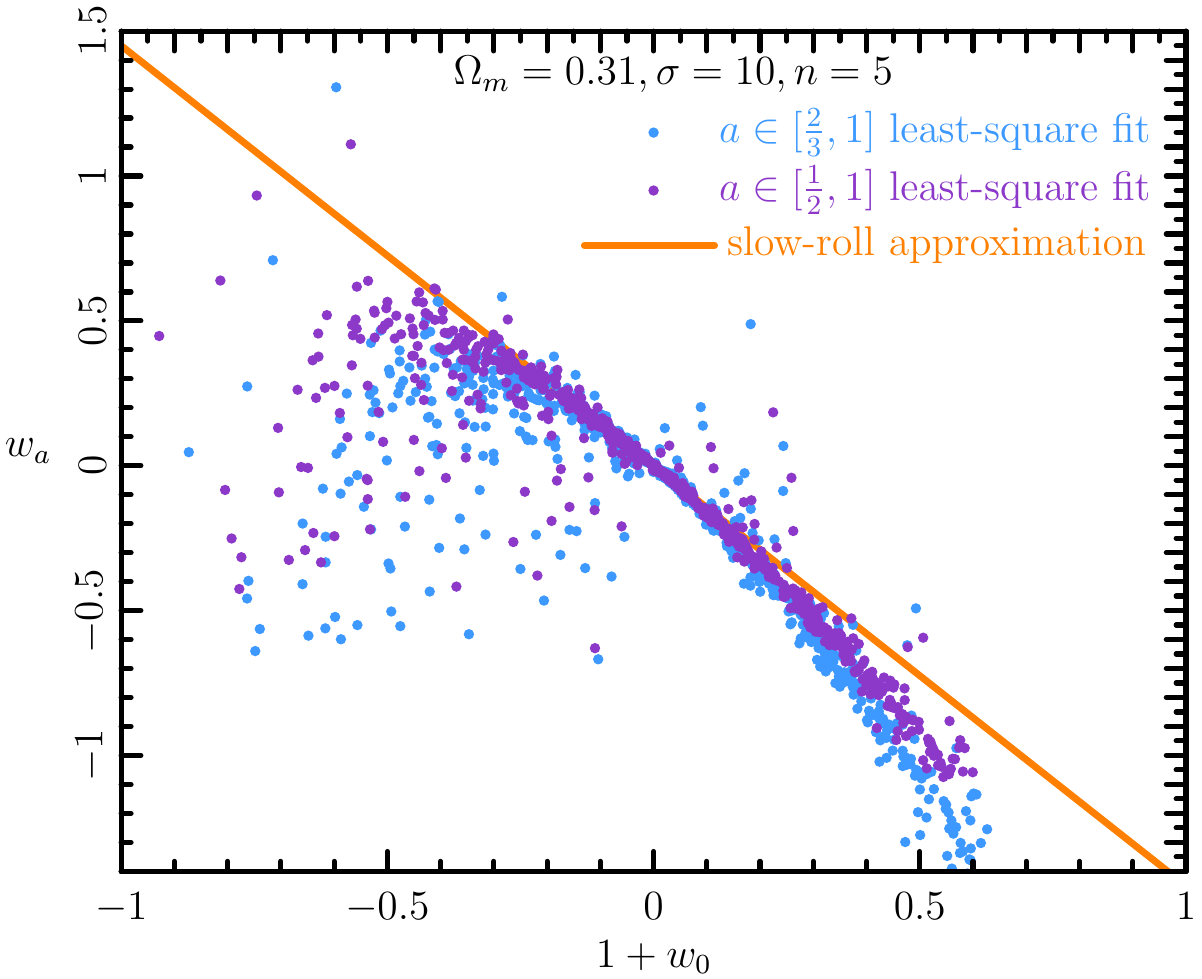}  \includegraphics[width=0.31\textwidth]{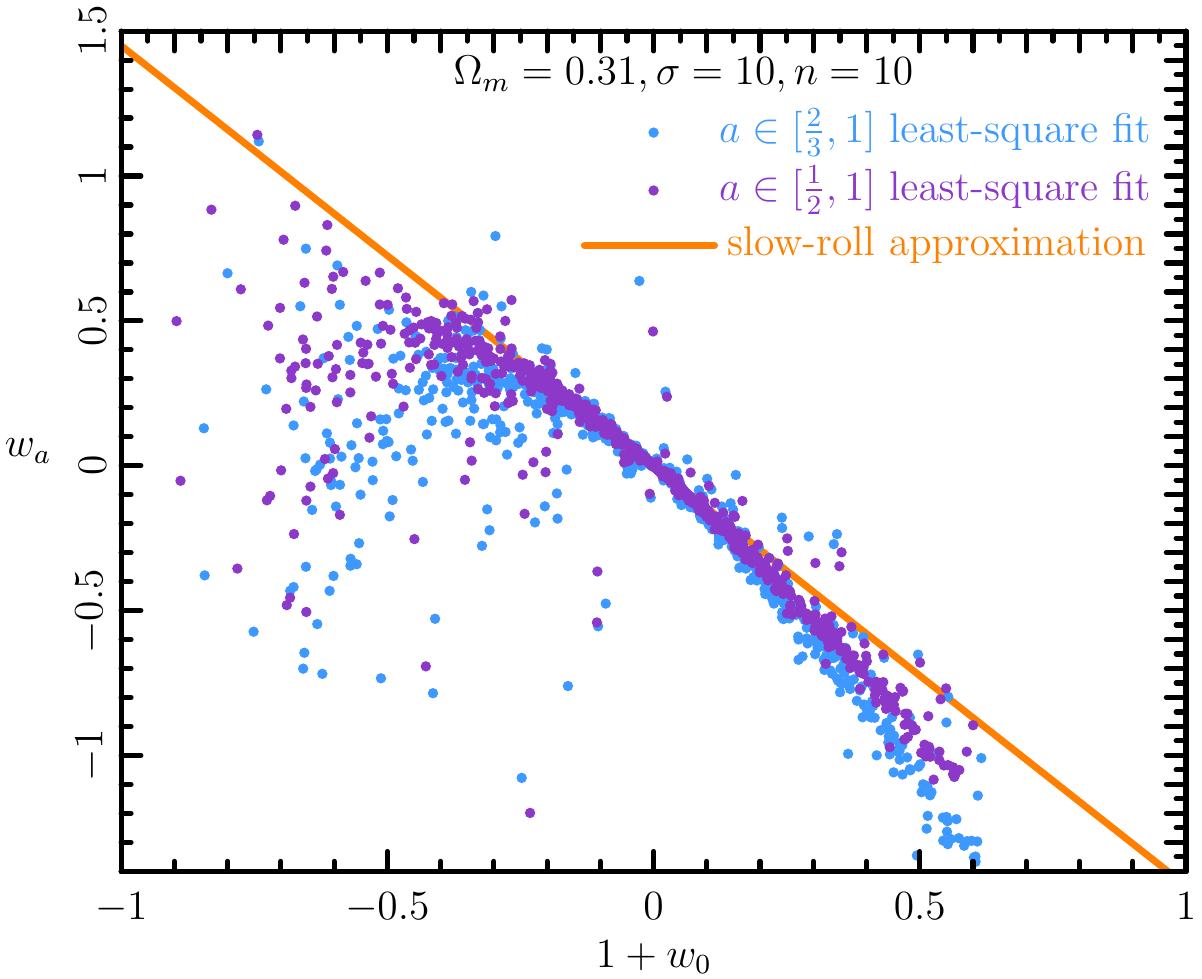}  \includegraphics[width=0.31\textwidth]{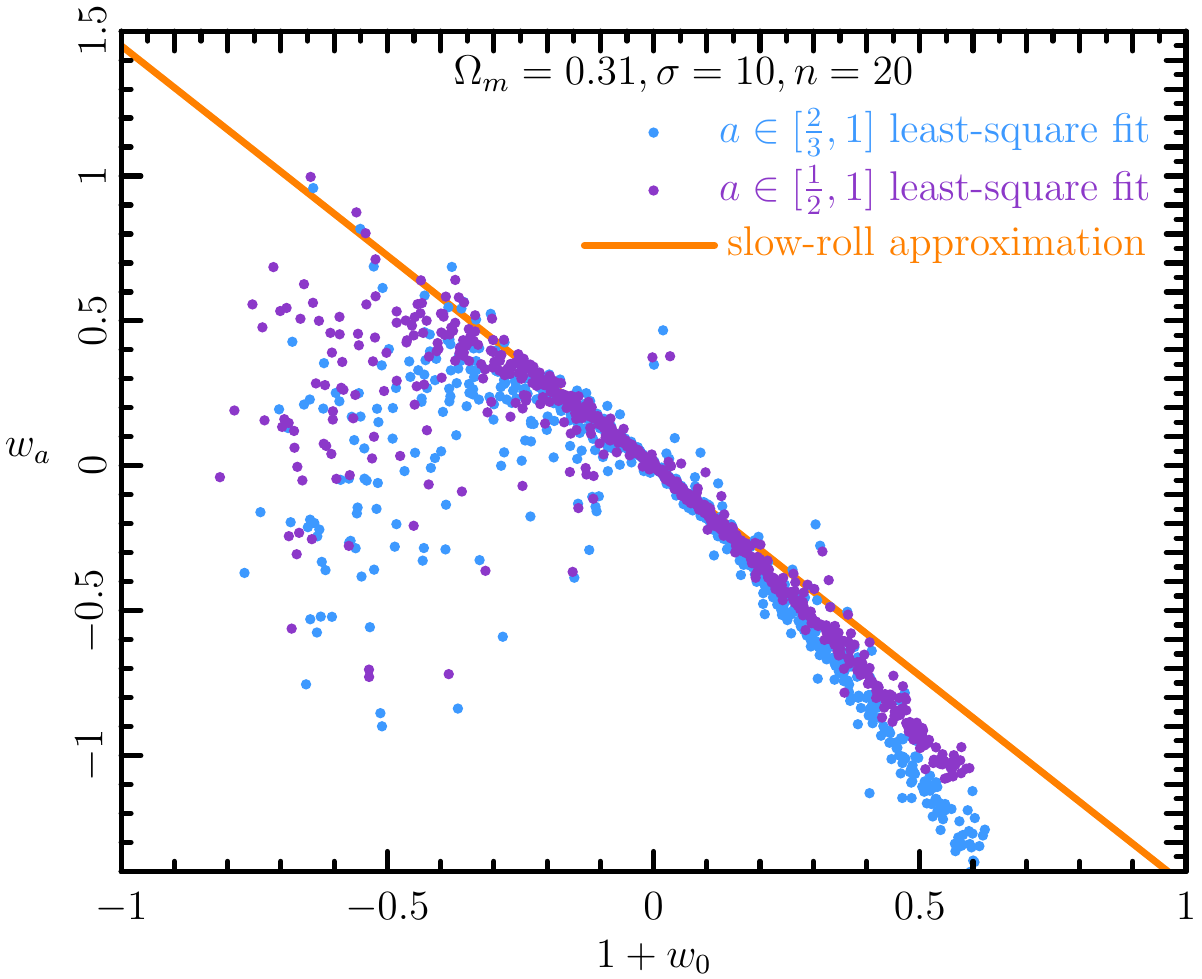}
  \caption{CPL fittings of randomly sampled thawing k-essence models. Each panel contains 1000 random samples. The orange line in each panel is the slow-roll approximation given by Eq.~\eqref{eq:slowroll}. \label{fig:w0wa}}
\end{figure*}

In Figure~\ref{fig:trajs} we plot some typical $w(a)$ trajectories for $(\sigma=3, n=10)$ and $(\sigma=10, n=20)$, respectively. We observe similar patterns of $w(a)$ trajectories in the two cases. The typical smoothness of $w(a)$ trajectories permits CPL fitting at low redshift, as we do and present in Figure~\ref{fig:w0wa}.
The similar distributions of $(w_0, w_a)$ under a variety of combination of sampling and linear-fitting parameters ($n=5, 10, 20$; $\sigma=1,5, 10$; $a_{\rm late}=\frac{1}{2}, \frac{2}{3}$) reveal that the qualitative pattern of $w(a)$ trajectories is independent of the sampling scheme. The result shows a universal clustering behavior in the $(w_0, w_a)$ space. About $\sim 90\%$  of the randomly generated thawing k-essence dark energy models cluster around a narrow band that is close to a slow-roll line, given by
\be
w_a\approx  -1.42 \left(\frac{\Omega_m}{0.3}\right)^{0.64}(1+w_0). \label{eq:slowroll}
\ee
\begin{figure}
  \includegraphics[width=\figwidth]{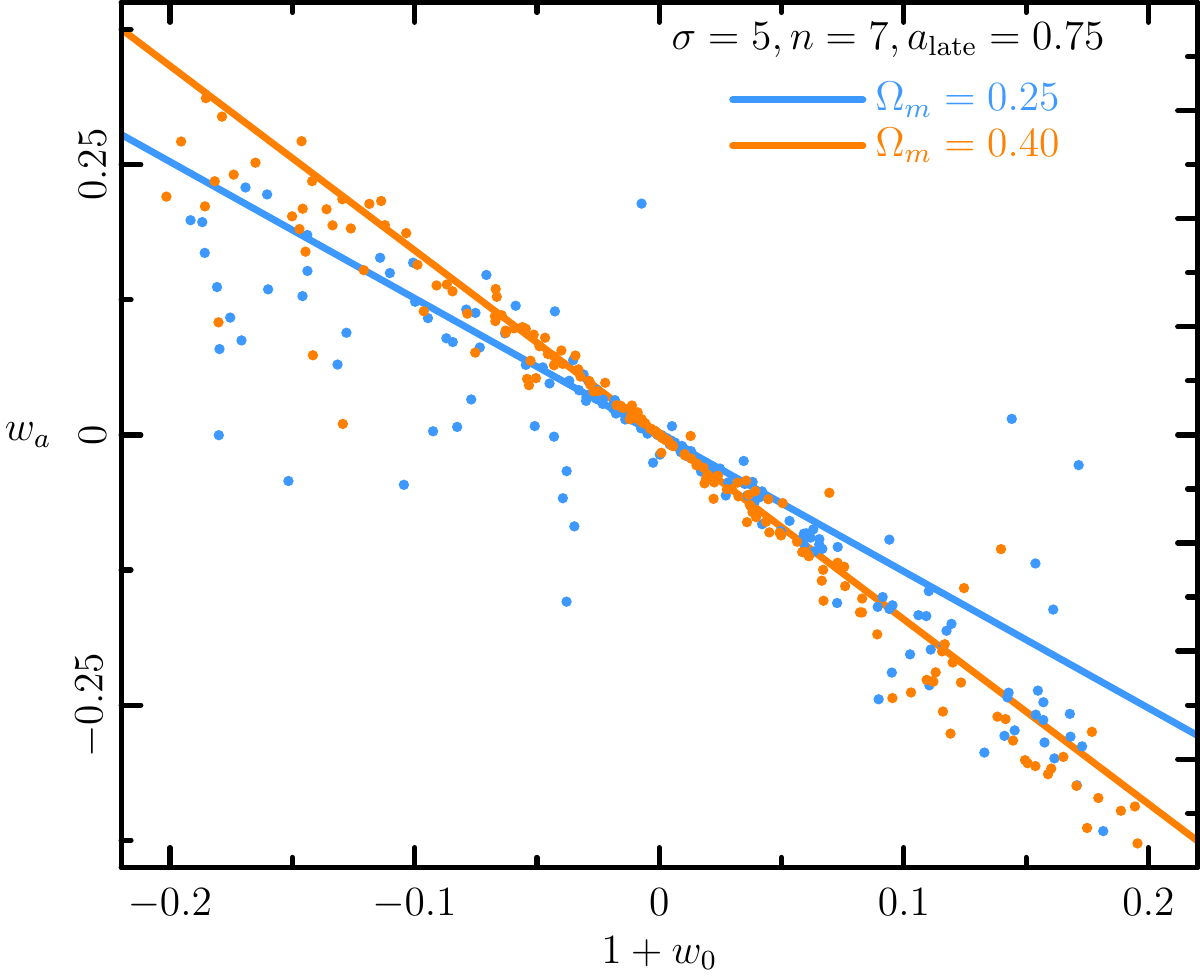}
  \caption{CPL fittings of randomly sampled thawing k-essence models with $|1+w|<0.2$, for $\Omega_m=0.25$ (200 skyblue dots) and $\Omega_m=0.4$ (200 orange dots) respectively. The solid lines are the slow-roll approximation given by Eq.~\eqref{eq:slowroll}. \label{fig:omm}}
\end{figure}
The dependence of the slow-roll line on $\Omega_m$ is demonstrated in Figure~\ref{fig:omm}, where we zoom out the slow-roll region by only keeping CPL fittings from models with $|1+w|<0.2$. 

Eq.~\eqref{eq:slowroll} can be understood semi-analytically. In the slow-roll limit, we may keep the low-est order terms in the Lagrangian $\mathcal{L} \approx V_{00} + V_{01}\phi + V_{10} X$. A redefinition of the field $\phi \rightarrow \frac{\phi}{\sqrt{\lvert V_{10}\rvert}}$ casts the model into a quintessence form $\mathcal{L}\approx X - f(\phi)$  (if $V_{10}<0$) or a phantom form  $\mathcal{L}\approx -X-f(\phi)$  (if $V_{10}>0$), where $f(\phi)=-V_{00}-\frac{V_{01}}{\sqrt{\lvert V_{10}\rvert}}\phi$. Ref.~\cite{WZ_Huang10} shows that in both cases $w(a)$ approximately has a functional form
\be
w(a) =   -1 + \frac{2\epsilon_s}{3}F^2\left[a\left(\frac{1-\Omega_m}{\Omega_m}\right)^{1/3}\right], \label{eq:wa}
\ee
where $\epsilon_s$ is a constant (for quintessence $\epsilon_s>0$ and for phantom $\epsilon_s<0$), and 
\be
F(x) \equiv  \frac{\sqrt{1+x^3}}{x^{3/2}}-\frac{\ln\left[x^{3/2}+\sqrt{1+x^3}\right]}{x^3}.
\ee
The unknown parameter $\epsilon_s$ cancels out in the ratio between $w_a \approx -\frac{dw}{da}\rvert_{a=1}$ and $1+w_0\approx 1+w\rvert_{a=1}$. We finally obtain
\be
\frac{w_a}{1+w_0} \approx \frac{-2\left(\frac{1-\Omega_m}{\Omega_m}\right)^{1/3}F'\left[\left(\frac{1-\Omega_m}{\Omega_m}\right)^{1/3}\right]}{F\left[\left(\frac{1-\Omega_m}{\Omega_m}\right)^{1/3}\right]}, \label{eq:srexact}
\ee
\begin{figure}
  \includegraphics[width=\figwidth]{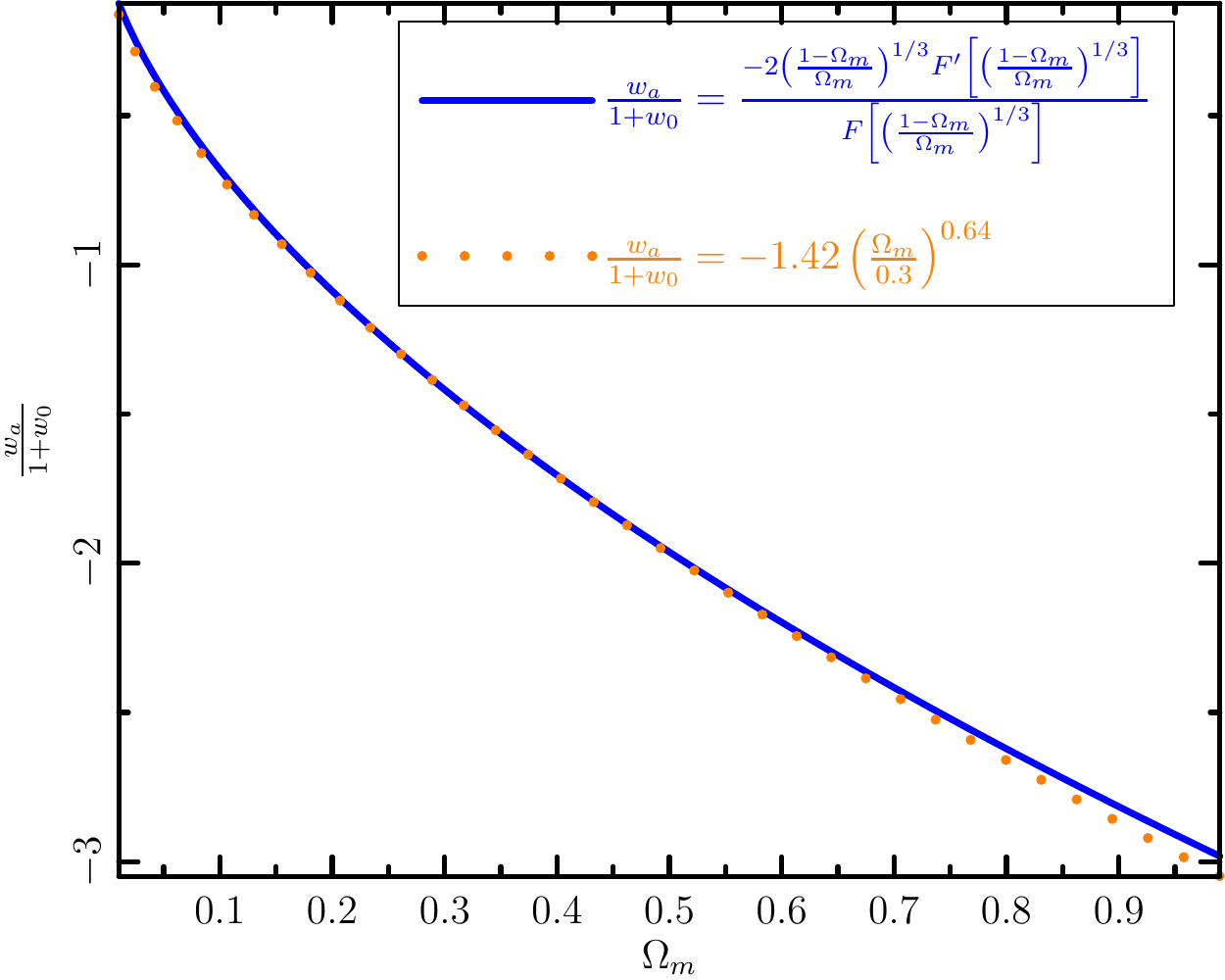}
  \caption{Comparison between Eq.~\eqref{eq:srexact} and its numeric approximation Eq.~\eqref{eq:slowroll}. \label{fig:wrat}}
\end{figure}
In Figure~\ref{fig:wrat} we show that Eq.~\eqref{eq:slowroll} is a good numeric approximation to Eq.~\eqref{eq:srexact}, especially in the proximity of $\Omega_m\sim 0.3$.

\section{Statistics of the Expansion Coefficients of Lagrangian}

\begin{figure*}
  \includegraphics[width=\textwidth]{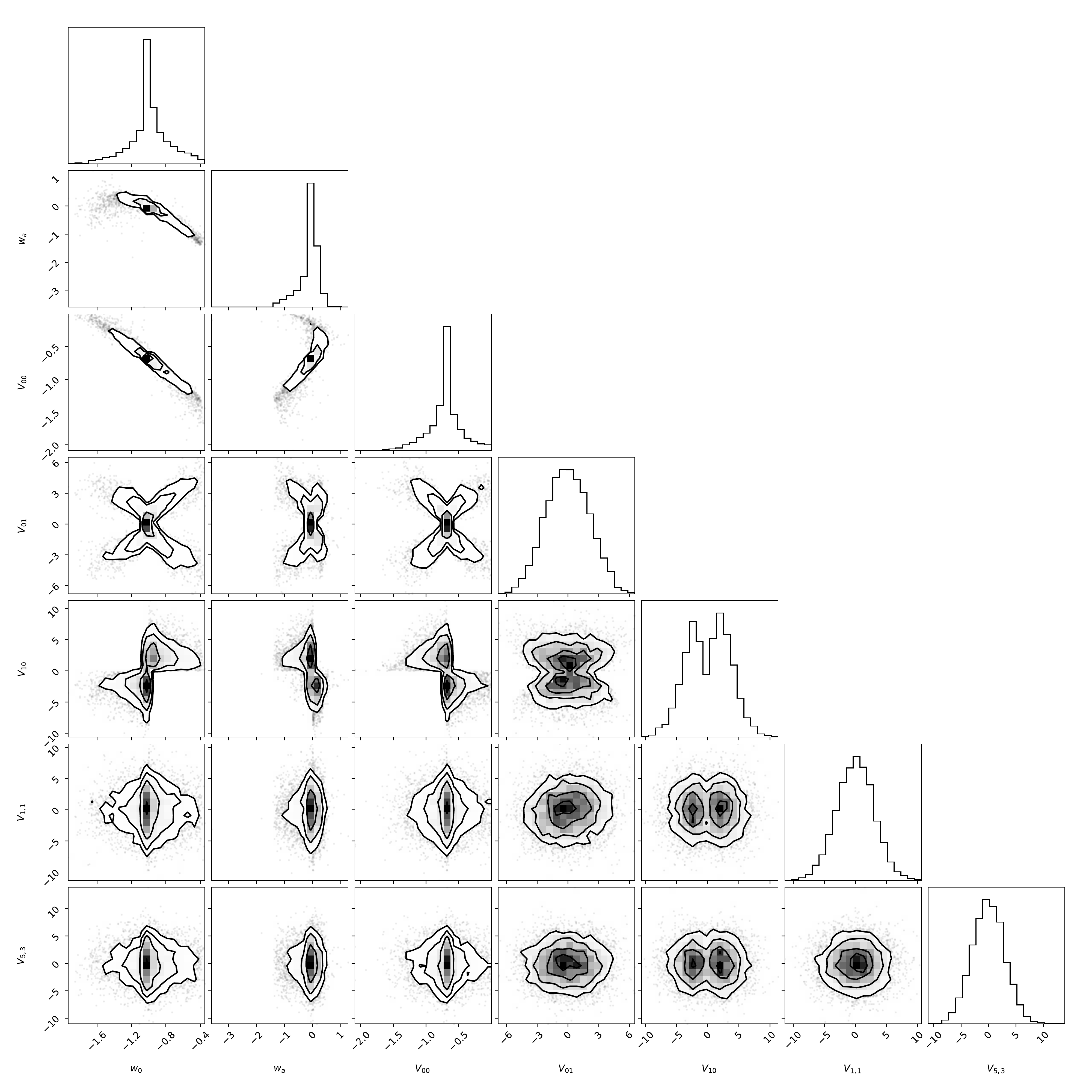}
  \caption{The marginalized $68.3\%$, $95.4\%$, $99.7\%$ confidence-level contours of $w_0$, $w_a$, $V_{00}$ and a few selected $V_{ij}$ coefficients. \label{fig:contours}}
\end{figure*}

We now turn our attention to the distribution of the $V_{ij}$ coefficients. Figure~\ref{fig:contours} shows the distribution of $w_0$, $w_a$, $V_{00}$, and a few selected $V_{ij}$ coefficients of 5000 thawing k-essence dark energy models generated with $\Omega_m=0.31, \sigma= 3, n=10, a_{\rm late} = 2/3$, and a broader scanning range $[-1000, -0.001]$ for $V_{00}$. 

When the present dark energy density $\rho_{\phi 0}=\Omega_\phi\rho_c$ is fixed, the early dark energy density $\rho_{\phi,\rm early} = -V_{00}\rho_c$ is strongly correlated with $w_0$ and $w_a$. If $\rho_{\phi,\rm early}>\rho_{\phi 0}$ (i.e. $V_{00}< -\Omega_\phi$), dark energy density tends to be a decreasing function of time, which corresponds to a quintessence-like solution ($1+w_0>0$). If $\rho_{\phi,\rm early}<\rho_{\phi 0}$ (i.e. $V_{00}> -\Omega_\phi$), dark energy density tends to be a increasing function of time, which corresponds to a phantom-like solution ($1+w_0<0$). Moreover, very few thawing k-essence dark energy models can be generated from $V_{00}\lesssim -2$ (i.e., $\rho_{\phi, \rm early} \gtrsim 3\rho_{\phi,\rm today}$), which typically corresponds to a fast-roll scenario that either violates the thawing assumption or fails to produce cosmic acceleration. Consequently, the scanning range of $V_{00}$, $[-100, -0.01]$ or a broader one, does not have a significant impact on our results.

The amplitude of $|V_{01}|$ gives the leading contribution to the driving force $S(\phi, X)$ in Eq.~\eqref{eq:bg}. Statistically speaking, a larger driving force ($\sim |V_{01}|$) leads to a larger kinetic energy ($\sim |1+w_0|$), making the butterfly shape distribution of $(w_0, V_{01})$ in Figure~\ref{fig:contours}.

The sign of $V_{10}$ determines the sign of the leading kinetic term in the Lagrangian. Therefore, a positive $V_{10}$ typically leads to a quintessence-like solution where $1+w_0>0$, and a negative $V_{10}$ typically leads to a phantom-like solution where $1+w_0<0$.  These features can be clearly seen in the $(w_0, V_{10})$ panel of Figure~\ref{fig:contours}.

Our approach does not resolve the fine-tuning problem of dark energy, because we have already used the unnaturally small $\rho_c$ as a unit to expand the Lagrangian density of k-essence.  The typical $\sim O(1)$ values of $V_{ij}$ parameters in Figure~\ref{fig:contours} indicate that our approach does not require additional fine-tuning beyond the usual one.

\section{Discussion and Conclusions}

The CPL parameterization \eqref{eq:w0wa} is used in the standard measure of figure of merit for future dark energy surveys. However, dark energy models are degenerate in the $(w_0, w_a)$ space. For example, in certain dynamical regimes an equivalence can be made between k-essence and quintessence~\cite{KandQ}. Non-clustering dark energy models with similar or identical homogeneous evolution may predict different horizon-scale perturbations, which, however  due to cosmic variance, are difficult to observationaly distinguish. In this work we seek a novel way to improve the statistics of model selection. Imagine that if $(w_0, w_a)$ measured by future surveys is far away from the clustering band shown in Figure~\ref{fig:w0wa}, thawing k-essence models would be disfavored at $\sim 90\%$ confidence level, and further including information from horizon-scale perturbations may allow to reject thawing k-essence models.

For technical reasons (e.g. the algorithm to solve $V_{00}$) we restrict the study in the thawing scenario. A thorough investigation of more complicated scenarios and many other dark energy models, and forecasts of dark-energy model selection for planned future surveys, would all be good exercises beyond the scope of this work.

\begin{figure}
  \includegraphics[width=\figwidth]{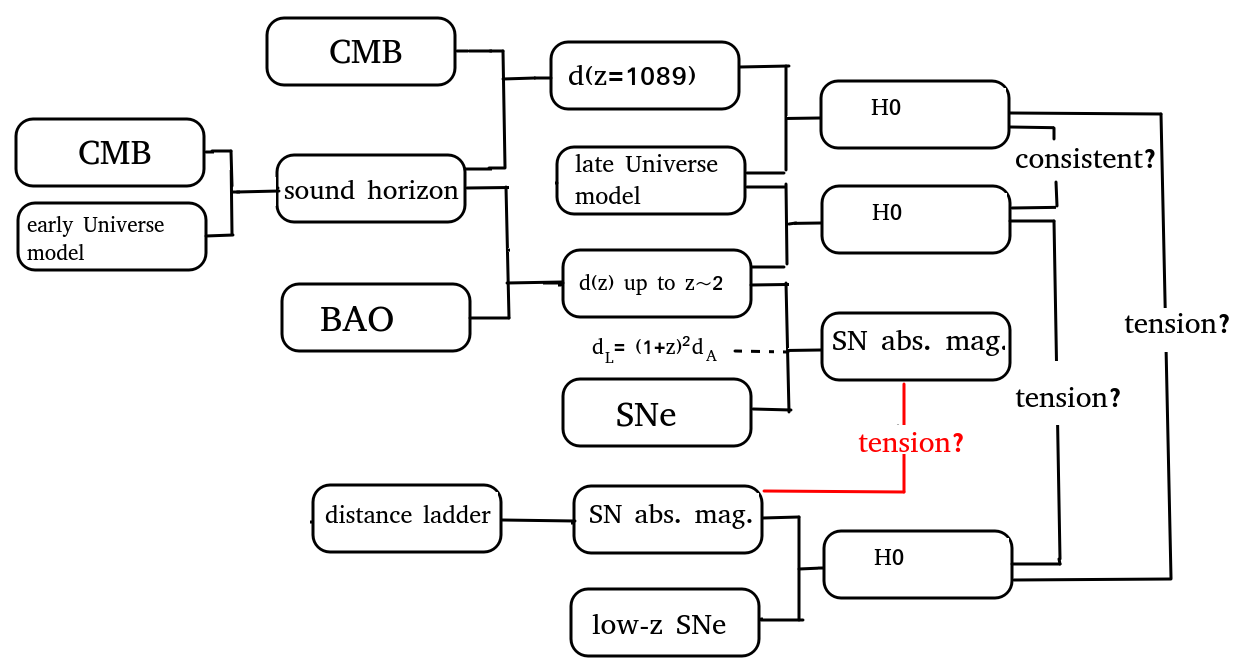}
  \caption{A different view of the Hubble tension~\cite{Efstathiou21}: the tension between the supernova absolute magnitude inferred from inverse distance ladder and that from the local distance ladder (the tension indicated by the red line) does not involve modeling of the late universe. \label{fig:H0}}
\end{figure}

As Figure~\ref{fig:trajs} shows, phantom-like ($w<-1$) solutions naturally arise in the randomly generated thawing k-essence models, which may ease the tension between CMB and distance-ladder measurements  of the Hubble constant~\cite{Huang16, Alam17, WZ_Miao18, Vagonzzi20, Alestas20, Bag21}. However, when BAO data are included, late-universe phantom-like models, with or without the prior $w_0$-$w_a$ distribution discussed in this work, cannot resolve the Hubble tension. This is because, as Ref.~\cite{Efstathiou21} pointed out and diagramatically shown in Figure~\ref{fig:H0},  Hubble tension is essentially a tension between the supernova absolute magnitude inferred from inverse distance ladder~\cite{Taubenberger2019, Lemos19} and that from the local distance ladder, which does not involve the modeling of the late universe.

\section{Acknowledgements}

This work is supported by the National key R\&D Program of China (Grant No. 2020YFC2201600), National Natural Science Foundation of China (NSFC) under Grant No. 12073088, National SKA Program of China No. 2020SKA0110402, and Guangdong Major Project of Basic and Applied Basic Research (Grant No. 2019B030302001). I also owe thanks to Prof. J. Richard Bond and Prof. Lev Kofman for intriguing discussion on this topic.


\begin{thebibliography}{29}%
\makeatletter
\providecommand \@ifxundefined [1]{%
 \@ifx{#1\undefined}
}%
\providecommand \@ifnum [1]{%
 \ifnum #1\expandafter \@firstoftwo
 \else \expandafter \@secondoftwo
 \fi
}%
\providecommand \@ifx [1]{%
 \ifx #1\expandafter \@firstoftwo
 \else \expandafter \@secondoftwo
 \fi
}%
\providecommand \natexlab [1]{#1}%
\providecommand \enquote  [1]{``#1''}%
\providecommand \bibnamefont  [1]{#1}%
\providecommand \bibfnamefont [1]{#1}%
\providecommand \citenamefont [1]{#1}%
\providecommand \href@noop [0]{\@secondoftwo}%
\providecommand \href [0]{\begingroup \@sanitize@url \@href}%
\providecommand \@href[1]{\@@startlink{#1}\@@href}%
\providecommand \@@href[1]{\endgroup#1\@@endlink}%
\providecommand \@sanitize@url [0]{\catcode `\\12\catcode `\$12\catcode
  `\&12\catcode `\#12\catcode `\^12\catcode `\_12\catcode `\%12\relax}%
\providecommand \@@startlink[1]{}%
\providecommand \@@endlink[0]{}%
\providecommand \url  [0]{\begingroup\@sanitize@url \@url }%
\providecommand \@url [1]{\endgroup\@href {#1}{\urlprefix }}%
\providecommand \urlprefix  [0]{URL }%
\providecommand \Eprint [0]{\href }%
\providecommand \doibase [0]{http://dx.doi.org/}%
\providecommand \selectlanguage [0]{\@gobble}%
\providecommand \bibinfo  [0]{\@secondoftwo}%
\providecommand \bibfield  [0]{\@secondoftwo}%
\providecommand \translation [1]{[#1]}%
\providecommand \BibitemOpen [0]{}%
\providecommand \bibitemStop [0]{}%
\providecommand \bibitemNoStop [0]{.\EOS\space}%
\providecommand \EOS [0]{\spacefactor3000\relax}%
\providecommand \BibitemShut  [1]{\csname bibitem#1\endcsname}%
\let\auto@bib@innerbib\@empty
\bibitem [{\citenamefont {Aghanim}\ \emph {et~al.}(2020)\citenamefont {Aghanim}
  \emph {et~al.}}]{Planck18Params}%
  \BibitemOpen
  \bibfield  {author} {\bibinfo {author} {\bibfnamefont {N.}~\bibnamefont
  {Aghanim}} \emph {et~al.} (\bibinfo {collaboration} {Planck}),\ }\href
  {\doibase 10.1051/0004-6361/201833910} {\bibfield  {journal} {\bibinfo
  {journal} {Astron. Astrophys.}\ }\textbf {\bibinfo {volume} {641}},\ \bibinfo
  {pages} {A6} (\bibinfo {year} {2020})},\ \Eprint
  {http://arxiv.org/abs/1807.06209} {arXiv:1807.06209 [astro-ph.CO]}
  \BibitemShut {NoStop}%
\bibitem [{\citenamefont {{Alam}}\ \emph {et~al.}(2021)\citenamefont {{Alam}}
  \emph {et~al.}}]{BOSS2021}%
  \BibitemOpen
  \bibfield  {author} {\bibinfo {author} {\bibfnamefont {S.}~\bibnamefont
  {{Alam}}} \emph {et~al.},\ }\href {\doibase 10.1103/PhysRevD.103.083533}
  {\bibfield  {journal} {\bibinfo  {journal} {\prd}\ }\textbf {\bibinfo
  {volume} {103}},\ \bibinfo {eid} {083533} (\bibinfo {year} {2021})},\ \Eprint
  {http://arxiv.org/abs/2007.08991} {arXiv:2007.08991 [astro-ph.CO]}
  \BibitemShut {NoStop}%
\bibitem [{\citenamefont {Abbott}\ \emph {et~al.}(2021)\citenamefont {Abbott}
  \emph {et~al.}}]{DES3yr}%
  \BibitemOpen
  \bibfield  {author} {\bibinfo {author} {\bibfnamefont {T.~M.~C.}\
  \bibnamefont {Abbott}} \emph {et~al.} (\bibinfo {collaboration} {DES}),\
  }\href@noop {} {\  (\bibinfo {year} {2021})},\ \Eprint
  {http://arxiv.org/abs/2105.13549} {arXiv:2105.13549 [astro-ph.CO]}
  \BibitemShut {NoStop}%
\bibitem [{\citenamefont {Scolnic}\ \emph {et~al.}(2018)\citenamefont {Scolnic}
  \emph {et~al.}}]{Pantheon}%
  \BibitemOpen
  \bibfield  {author} {\bibinfo {author} {\bibfnamefont {D.~M.}\ \bibnamefont
  {Scolnic}} \emph {et~al.},\ }\href {\doibase 10.3847/1538-4357/aab9bb}
  {\bibfield  {journal} {\bibinfo  {journal} {Astrophys. J.}\ }\textbf
  {\bibinfo {volume} {859}},\ \bibinfo {pages} {101} (\bibinfo {year}
  {2018})},\ \Eprint {http://arxiv.org/abs/1710.00845} {arXiv:1710.00845
  [astro-ph.CO]} \BibitemShut {NoStop}%
\bibitem [{\citenamefont {Riess}\ \emph {et~al.}(2021)\citenamefont {Riess},
  \citenamefont {Casertano}, \citenamefont {Yuan}, \citenamefont {Bowers},
  \citenamefont {Macri}, \citenamefont {Zinn},\ and\ \citenamefont
  {Scolnic}}]{Riess21}%
  \BibitemOpen
  \bibfield  {author} {\bibinfo {author} {\bibfnamefont {A.~G.}\ \bibnamefont
  {Riess}}, \bibinfo {author} {\bibfnamefont {S.}~\bibnamefont {Casertano}},
  \bibinfo {author} {\bibfnamefont {W.}~\bibnamefont {Yuan}}, \bibinfo {author}
  {\bibfnamefont {J.~B.}\ \bibnamefont {Bowers}}, \bibinfo {author}
  {\bibfnamefont {L.}~\bibnamefont {Macri}}, \bibinfo {author} {\bibfnamefont
  {J.~C.}\ \bibnamefont {Zinn}}, \ and\ \bibinfo {author} {\bibfnamefont
  {D.}~\bibnamefont {Scolnic}},\ }\href {\doibase 10.3847/2041-8213/abdbaf}
  {\bibfield  {journal} {\bibinfo  {journal} {Astrophys. J. Lett.}\ }\textbf
  {\bibinfo {volume} {908}},\ \bibinfo {pages} {L6} (\bibinfo {year} {2021})},\
  \Eprint {http://arxiv.org/abs/2012.08534} {arXiv:2012.08534 [astro-ph.CO]}
  \BibitemShut {NoStop}%
\bibitem [{\citenamefont {{Ostrogradsky}}(1850)}]{Ostrogradsky}%
  \BibitemOpen
  \bibfield  {author} {\bibinfo {author} {\bibfnamefont {M.}~\bibnamefont
  {{Ostrogradsky}}},\ }\href@noop {} {\bibfield  {journal} {\bibinfo  {journal}
  {Mem. Acad. St. Petersbourg}\ }\textbf {\bibinfo {volume} {series 6}},\
  \bibinfo {pages} {385} (\bibinfo {year} {1850})}\BibitemShut {NoStop}%
\bibitem [{\citenamefont {{Armendariz-Picon}}\ \emph
  {et~al.}(2000)\citenamefont {{Armendariz-Picon}}, \citenamefont
  {{Mukhanov}},\ and\ \citenamefont {{Steinhardt}}}]{Kessence_APC}%
  \BibitemOpen
  \bibfield  {author} {\bibinfo {author} {\bibfnamefont {C.}~\bibnamefont
  {{Armendariz-Picon}}}, \bibinfo {author} {\bibfnamefont {V.}~\bibnamefont
  {{Mukhanov}}}, \ and\ \bibinfo {author} {\bibfnamefont {P.~J.}\ \bibnamefont
  {{Steinhardt}}},\ }\href {\doibase 10.1103/PhysRevLett.85.4438} {\bibfield
  {journal} {\bibinfo  {journal} {\prl}\ }\textbf {\bibinfo {volume} {85}},\
  \bibinfo {pages} {4438} (\bibinfo {year} {2000})},\ \Eprint
  {http://arxiv.org/abs/astro-ph/0004134} {arXiv:astro-ph/0004134 [astro-ph]}
  \BibitemShut {NoStop}%
\bibitem [{\citenamefont {{Malquarti}}\ \emph
  {et~al.}(2003{\natexlab{a}})\citenamefont {{Malquarti}}, \citenamefont
  {{Copeland}},\ and\ \citenamefont {{Liddle}}}]{Coincidence}%
  \BibitemOpen
  \bibfield  {author} {\bibinfo {author} {\bibfnamefont {M.}~\bibnamefont
  {{Malquarti}}}, \bibinfo {author} {\bibfnamefont {E.~J.}\ \bibnamefont
  {{Copeland}}}, \ and\ \bibinfo {author} {\bibfnamefont {A.~R.}\ \bibnamefont
  {{Liddle}}},\ }\href {\doibase 10.1103/PhysRevD.68.023512} {\bibfield
  {journal} {\bibinfo  {journal} {\prd}\ }\textbf {\bibinfo {volume} {68}},\
  \bibinfo {eid} {023512} (\bibinfo {year} {2003}{\natexlab{a}})},\ \Eprint
  {http://arxiv.org/abs/astro-ph/0304277} {arXiv:astro-ph/0304277 [astro-ph]}
  \BibitemShut {NoStop}%
\bibitem [{\citenamefont {Chevallier}\ and\ \citenamefont
  {Polarski}(2001)}]{Chevallier:2000qy}%
  \BibitemOpen
  \bibfield  {author} {\bibinfo {author} {\bibfnamefont {M.}~\bibnamefont
  {Chevallier}}\ and\ \bibinfo {author} {\bibfnamefont {D.}~\bibnamefont
  {Polarski}},\ }\href {\doibase 10.1142/S0218271801000822} {\bibfield
  {journal} {\bibinfo  {journal} {Int. J. Mod. Phys. D}\ }\textbf {\bibinfo
  {volume} {10}},\ \bibinfo {pages} {213} (\bibinfo {year} {2001})},\ \Eprint
  {http://arxiv.org/abs/gr-qc/0009008} {arXiv:gr-qc/0009008} \BibitemShut
  {NoStop}%
\bibitem [{\citenamefont {Linder}(2003)}]{Linder:2002et}%
  \BibitemOpen
  \bibfield  {author} {\bibinfo {author} {\bibfnamefont {E.~V.}\ \bibnamefont
  {Linder}},\ }\href {\doibase 10.1103/PhysRevLett.90.091301} {\bibfield
  {journal} {\bibinfo  {journal} {Phys. Rev. Lett.}\ }\textbf {\bibinfo
  {volume} {90}},\ \bibinfo {pages} {091301} (\bibinfo {year} {2003})},\
  \Eprint {http://arxiv.org/abs/astro-ph/0208512} {arXiv:astro-ph/0208512}
  \BibitemShut {NoStop}%
\bibitem [{\citenamefont {{Crittenden}}\ \emph {et~al.}(2007)\citenamefont
  {{Crittenden}}, \citenamefont {{Majerotto}},\ and\ \citenamefont
  {{Piazza}}}]{Crittenden2007}%
  \BibitemOpen
  \bibfield  {author} {\bibinfo {author} {\bibfnamefont {R.}~\bibnamefont
  {{Crittenden}}}, \bibinfo {author} {\bibfnamefont {E.}~\bibnamefont
  {{Majerotto}}}, \ and\ \bibinfo {author} {\bibfnamefont {F.}~\bibnamefont
  {{Piazza}}},\ }\href {\doibase 10.1103/PhysRevLett.98.251301} {\bibfield
  {journal} {\bibinfo  {journal} {Physical Review Letters}\ }\textbf {\bibinfo
  {volume} {98}},\ \bibinfo {pages} {251301} (\bibinfo {year} {2007})},\
  \Eprint {http://arxiv.org/abs/arXiv:astro-ph/0702003}
  {arXiv:astro-ph/0702003} \BibitemShut {NoStop}%
\bibitem [{\citenamefont {{Scherrer}}\ and\ \citenamefont
  {{Sen}}(2008)}]{Scherrer2008}%
  \BibitemOpen
  \bibfield  {author} {\bibinfo {author} {\bibfnamefont {R.~J.}\ \bibnamefont
  {{Scherrer}}}\ and\ \bibinfo {author} {\bibfnamefont {A.~A.}\ \bibnamefont
  {{Sen}}},\ }\href {\doibase 10.1103/PhysRevD.77.083515} {\bibfield  {journal}
  {\bibinfo  {journal} {\prd}\ }\textbf {\bibinfo {volume} {77}},\ \bibinfo
  {pages} {083515} (\bibinfo {year} {2008})},\ \Eprint
  {http://arxiv.org/abs/arXiv:0712.3450} {arXiv:0712.3450} \BibitemShut
  {NoStop}%
\bibitem [{\citenamefont {{Chiba}}(2009)}]{Chiba2009}%
  \BibitemOpen
  \bibfield  {author} {\bibinfo {author} {\bibfnamefont {T.}~\bibnamefont
  {{Chiba}}},\ }\href@noop {} {\bibfield  {journal} {\bibinfo  {journal}
  {\prd}\ }\textbf {\bibinfo {volume} {79}},\ \bibinfo {pages} {083517}
  (\bibinfo {year} {2009})},\ \Eprint {http://arxiv.org/abs/0902.4037}
  {arXiv:0902.4037} \BibitemShut {NoStop}%
\bibitem [{\citenamefont {Zhiqi~Huang}(2011)}]{WZ_Huang10}%
  \BibitemOpen
  \bibfield  {author} {\bibinfo {author} {\bibfnamefont {L.~K.}\ \bibnamefont
  {Zhiqi~Huang}, \bibfnamefont {J.~Richard~Bond}},\ }\href {\doibase
  10.1088/0004-637X/726/2/64} {\bibfield  {journal} {\bibinfo  {journal}
  {{ApJ}}\ }\textbf {\bibinfo {volume} {726}},\ \bibinfo {pages} {64} (\bibinfo
  {year} {2011})},\ \Eprint {http://arxiv.org/abs/1007.5297} {arXiv:1007.5297}
  \BibitemShut {NoStop}%
\bibitem [{\citenamefont {Haitao~Miao}(2018)}]{WZ_Miao18}%
  \BibitemOpen
  \bibfield  {author} {\bibinfo {author} {\bibfnamefont {Z.~H.}\ \bibnamefont
  {Haitao~Miao}},\ }\href {\doibase 10.3847/1538-4357/aae523} {\bibfield
  {journal} {\bibinfo  {journal} {{ApJ}}\ }\textbf {\bibinfo {volume} {868}},\
  \bibinfo {pages} {20} (\bibinfo {year} {2018})},\ \Eprint
  {http://arxiv.org/abs/1803.07320} {arXiv:1803.07320} \BibitemShut {NoStop}%
\bibitem [{\citenamefont {Chiba}\ \emph {et~al.}(2009)\citenamefont {Chiba},
  \citenamefont {Dutta},\ and\ \citenamefont {Scherrer}}]{WZ_Chiba}%
  \BibitemOpen
  \bibfield  {author} {\bibinfo {author} {\bibfnamefont {T.}~\bibnamefont
  {Chiba}}, \bibinfo {author} {\bibfnamefont {S.}~\bibnamefont {Dutta}}, \ and\
  \bibinfo {author} {\bibfnamefont {R.~J.}\ \bibnamefont {Scherrer}},\ }\href
  {\doibase 10.1103/PhysRevD.80.043517} {\bibfield  {journal} {\bibinfo
  {journal} {Phys. Rev. D}\ }\textbf {\bibinfo {volume} {80}},\ \bibinfo
  {pages} {043517} (\bibinfo {year} {2009})},\ \Eprint
  {http://arxiv.org/abs/0906.0628} {arXiv:0906.0628 [astro-ph.CO]} \BibitemShut
  {NoStop}%
\bibitem [{\citenamefont {{Kehayias}}\ and\ \citenamefont
  {{Scherrer}}(2019)}]{WZ_Kehayias}%
  \BibitemOpen
  \bibfield  {author} {\bibinfo {author} {\bibfnamefont {J.}~\bibnamefont
  {{Kehayias}}}\ and\ \bibinfo {author} {\bibfnamefont {R.~J.}\ \bibnamefont
  {{Scherrer}}},\ }\href {\doibase 10.1103/PhysRevD.100.023525} {\bibfield
  {journal} {\bibinfo  {journal} {\prd}\ }\textbf {\bibinfo {volume} {100}},\
  \bibinfo {eid} {023525} (\bibinfo {year} {2019})},\ \Eprint
  {http://arxiv.org/abs/1905.05628} {arXiv:1905.05628 [gr-qc]} \BibitemShut
  {NoStop}%
\bibitem [{\citenamefont {Huterer}\ and\ \citenamefont
  {Peiris}(2007)}]{HP2007}%
  \BibitemOpen
  \bibfield  {author} {\bibinfo {author} {\bibfnamefont {D.}~\bibnamefont
  {Huterer}}\ and\ \bibinfo {author} {\bibfnamefont {H.~V.}\ \bibnamefont
  {Peiris}},\ }\href {\doibase 10.1103/PhysRevD.75.083503} {\bibfield
  {journal} {\bibinfo  {journal} {Phys. Rev. D}\ }\textbf {\bibinfo {volume}
  {75}},\ \bibinfo {pages} {083503} (\bibinfo {year} {2007})},\ \Eprint
  {http://arxiv.org/abs/astro-ph/0610427} {arXiv:astro-ph/0610427} \BibitemShut
  {NoStop}%
\bibitem [{\citenamefont {{Marsh}}\ \emph {et~al.}(2014)\citenamefont
  {{Marsh}}, \citenamefont {{Bull}}, \citenamefont {{Ferreira}},\ and\
  \citenamefont {{Pontzen}}}]{Marsh14}%
  \BibitemOpen
  \bibfield  {author} {\bibinfo {author} {\bibfnamefont {D.~J.~E.}\
  \bibnamefont {{Marsh}}}, \bibinfo {author} {\bibfnamefont {P.}~\bibnamefont
  {{Bull}}}, \bibinfo {author} {\bibfnamefont {P.~G.}\ \bibnamefont
  {{Ferreira}}}, \ and\ \bibinfo {author} {\bibfnamefont {A.}~\bibnamefont
  {{Pontzen}}},\ }\href {\doibase 10.1103/PhysRevD.90.105023} {\bibfield
  {journal} {\bibinfo  {journal} {\prd}\ }\textbf {\bibinfo {volume} {90}},\
  \bibinfo {eid} {105023} (\bibinfo {year} {2014})},\ \Eprint
  {http://arxiv.org/abs/1406.2301} {arXiv:1406.2301 [astro-ph.CO]} \BibitemShut
  {NoStop}%
\bibitem [{\citenamefont {Garc\'\i{}a-Garc\'\i{}a}\ \emph
  {et~al.}(2020)\citenamefont {Garc\'\i{}a-Garc\'\i{}a}, \citenamefont
  {Bellini}, \citenamefont {Ferreira}, \citenamefont {Traykova},\ and\
  \citenamefont {Zumalac\'arregui}}]{Garcia19}%
  \BibitemOpen
  \bibfield  {author} {\bibinfo {author} {\bibfnamefont {C.}~\bibnamefont
  {Garc\'\i{}a-Garc\'\i{}a}}, \bibinfo {author} {\bibfnamefont
  {E.}~\bibnamefont {Bellini}}, \bibinfo {author} {\bibfnamefont {P.~G.}\
  \bibnamefont {Ferreira}}, \bibinfo {author} {\bibfnamefont {D.}~\bibnamefont
  {Traykova}}, \ and\ \bibinfo {author} {\bibfnamefont {M.}~\bibnamefont
  {Zumalac\'arregui}},\ }\href {\doibase 10.1103/PhysRevD.101.063508}
  {\bibfield  {journal} {\bibinfo  {journal} {Phys. Rev. D}\ }\textbf {\bibinfo
  {volume} {101}},\ \bibinfo {pages} {063508} (\bibinfo {year} {2020})},\
  \Eprint {http://arxiv.org/abs/1911.02868} {arXiv:1911.02868 [astro-ph.CO]}
  \BibitemShut {NoStop}%
\bibitem [{\citenamefont {{Malquarti}}\ \emph
  {et~al.}(2003{\natexlab{b}})\citenamefont {{Malquarti}}, \citenamefont
  {{Copeland}}, \citenamefont {{Liddle}},\ and\ \citenamefont
  {{Trodden}}}]{KandQ}%
  \BibitemOpen
  \bibfield  {author} {\bibinfo {author} {\bibfnamefont {M.}~\bibnamefont
  {{Malquarti}}}, \bibinfo {author} {\bibfnamefont {E.~J.}\ \bibnamefont
  {{Copeland}}}, \bibinfo {author} {\bibfnamefont {A.~R.}\ \bibnamefont
  {{Liddle}}}, \ and\ \bibinfo {author} {\bibfnamefont {M.}~\bibnamefont
  {{Trodden}}},\ }\href {\doibase 10.1103/PhysRevD.67.123503} {\bibfield
  {journal} {\bibinfo  {journal} {\prd}\ }\textbf {\bibinfo {volume} {67}},\
  \bibinfo {eid} {123503} (\bibinfo {year} {2003}{\natexlab{b}})},\ \Eprint
  {http://arxiv.org/abs/astro-ph/0302279} {arXiv:astro-ph/0302279 [astro-ph]}
  \BibitemShut {NoStop}%
\bibitem [{\citenamefont {{Huang}}\ and\ \citenamefont
  {{Wang}}(2016)}]{Huang16}%
  \BibitemOpen
  \bibfield  {author} {\bibinfo {author} {\bibfnamefont {Q.-G.}\ \bibnamefont
  {{Huang}}}\ and\ \bibinfo {author} {\bibfnamefont {K.}~\bibnamefont
  {{Wang}}},\ }\href {\doibase 10.1140/epjc/s10052-016-4352-x} {\bibfield
  {journal} {\bibinfo  {journal} {European Physical Journal C}\ }\textbf
  {\bibinfo {volume} {76}},\ \bibinfo {eid} {506} (\bibinfo {year} {2016})},\
  \Eprint {http://arxiv.org/abs/1606.05965} {arXiv:1606.05965 [astro-ph.CO]}
  \BibitemShut {NoStop}%
\bibitem [{\citenamefont {{Alam}}\ \emph {et~al.}(2017)\citenamefont {{Alam}},
  \citenamefont {{Bag}},\ and\ \citenamefont {{Sahni}}}]{Alam17}%
  \BibitemOpen
  \bibfield  {author} {\bibinfo {author} {\bibfnamefont {U.}~\bibnamefont
  {{Alam}}}, \bibinfo {author} {\bibfnamefont {S.}~\bibnamefont {{Bag}}}, \
  and\ \bibinfo {author} {\bibfnamefont {V.}~\bibnamefont {{Sahni}}},\ }\href
  {\doibase 10.1103/PhysRevD.95.023524} {\bibfield  {journal} {\bibinfo
  {journal} {\prd}\ }\textbf {\bibinfo {volume} {95}},\ \bibinfo {eid} {023524}
  (\bibinfo {year} {2017})},\ \Eprint {http://arxiv.org/abs/1605.04707}
  {arXiv:1605.04707 [astro-ph.CO]} \BibitemShut {NoStop}%
\bibitem [{\citenamefont {{Vagnozzi}}(2020)}]{Vagonzzi20}%
  \BibitemOpen
  \bibfield  {author} {\bibinfo {author} {\bibfnamefont {S.}~\bibnamefont
  {{Vagnozzi}}},\ }\href {\doibase 10.1103/PhysRevD.102.023518} {\bibfield
  {journal} {\bibinfo  {journal} {\prd}\ }\textbf {\bibinfo {volume} {102}},\
  \bibinfo {eid} {023518} (\bibinfo {year} {2020})},\ \Eprint
  {http://arxiv.org/abs/1907.07569} {arXiv:1907.07569 [astro-ph.CO]}
  \BibitemShut {NoStop}%
\bibitem [{\citenamefont {{Alestas}}\ \emph {et~al.}(2020)\citenamefont
  {{Alestas}}, \citenamefont {{Kazantzidis}},\ and\ \citenamefont
  {{Perivolaropoulos}}}]{Alestas20}%
  \BibitemOpen
  \bibfield  {author} {\bibinfo {author} {\bibfnamefont {G.}~\bibnamefont
  {{Alestas}}}, \bibinfo {author} {\bibfnamefont {L.}~\bibnamefont
  {{Kazantzidis}}}, \ and\ \bibinfo {author} {\bibfnamefont {L.}~\bibnamefont
  {{Perivolaropoulos}}},\ }\href {\doibase 10.1103/PhysRevD.101.123516}
  {\bibfield  {journal} {\bibinfo  {journal} {\prd}\ }\textbf {\bibinfo
  {volume} {101}},\ \bibinfo {eid} {123516} (\bibinfo {year} {2020})},\ \Eprint
  {http://arxiv.org/abs/2004.08363} {arXiv:2004.08363 [astro-ph.CO]}
  \BibitemShut {NoStop}%
\bibitem [{\citenamefont {{Bag}}\ \emph {et~al.}(2021)\citenamefont {{Bag}},
  \citenamefont {{Sahni}}, \citenamefont {{Shafieloo}},\ and\ \citenamefont
  {{Shtanov}}}]{Bag21}%
  \BibitemOpen
  \bibfield  {author} {\bibinfo {author} {\bibfnamefont {S.}~\bibnamefont
  {{Bag}}}, \bibinfo {author} {\bibfnamefont {V.}~\bibnamefont {{Sahni}}},
  \bibinfo {author} {\bibfnamefont {A.}~\bibnamefont {{Shafieloo}}}, \ and\
  \bibinfo {author} {\bibfnamefont {Y.}~\bibnamefont {{Shtanov}}},\ }\href@noop
  {} {\bibfield  {journal} {\bibinfo  {journal} {arXiv e-prints}\ ,\ \bibinfo
  {eid} {arXiv:2107.03271}} (\bibinfo {year} {2021})},\ \Eprint
  {http://arxiv.org/abs/2107.03271} {arXiv:2107.03271 [astro-ph.CO]}
  \BibitemShut {NoStop}%
\bibitem [{\citenamefont {{Efstathiou}}(2021)}]{Efstathiou21}%
  \BibitemOpen
  \bibfield  {author} {\bibinfo {author} {\bibfnamefont {G.}~\bibnamefont
  {{Efstathiou}}},\ }\href {\doibase 10.1093/mnras/stab1588} {\bibfield
  {journal} {\bibinfo  {journal} {\mnras}\ }\textbf {\bibinfo {volume} {505}},\
  \bibinfo {pages} {3866} (\bibinfo {year} {2021})},\ \Eprint
  {http://arxiv.org/abs/2103.08723} {arXiv:2103.08723 [astro-ph.CO]}
  \BibitemShut {NoStop}%
\bibitem [{\citenamefont {{Taubenberger}}\ \emph {et~al.}(2019)\citenamefont
  {{Taubenberger}}, \citenamefont {{Suyu}}, \citenamefont {{Komatsu}},
  \citenamefont {{Jee}}, \citenamefont {{Birrer}}, \citenamefont {{Bonvin}},
  \citenamefont {{Courbin}}, \citenamefont {{Rusu}}, \citenamefont {{Shajib}},\
  and\ \citenamefont {{Wong}}}]{Taubenberger2019}%
  \BibitemOpen
  \bibfield  {author} {\bibinfo {author} {\bibfnamefont {S.}~\bibnamefont
  {{Taubenberger}}}, \bibinfo {author} {\bibfnamefont {S.~H.}\ \bibnamefont
  {{Suyu}}}, \bibinfo {author} {\bibfnamefont {E.}~\bibnamefont {{Komatsu}}},
  \bibinfo {author} {\bibfnamefont {I.}~\bibnamefont {{Jee}}}, \bibinfo
  {author} {\bibfnamefont {S.}~\bibnamefont {{Birrer}}}, \bibinfo {author}
  {\bibfnamefont {V.}~\bibnamefont {{Bonvin}}}, \bibinfo {author}
  {\bibfnamefont {F.}~\bibnamefont {{Courbin}}}, \bibinfo {author}
  {\bibfnamefont {C.~E.}\ \bibnamefont {{Rusu}}}, \bibinfo {author}
  {\bibfnamefont {A.~J.}\ \bibnamefont {{Shajib}}}, \ and\ \bibinfo {author}
  {\bibfnamefont {K.~C.}\ \bibnamefont {{Wong}}},\ }\href {\doibase
  10.1051/0004-6361/201935980} {\bibfield  {journal} {\bibinfo  {journal}
  {\aap}\ }\textbf {\bibinfo {volume} {628}},\ \bibinfo {eid} {L7} (\bibinfo
  {year} {2019})},\ \Eprint {http://arxiv.org/abs/1905.12496} {arXiv:1905.12496
  [astro-ph.CO]} \BibitemShut {NoStop}%
\bibitem [{\citenamefont {{Lemos}}\ \emph {et~al.}(2019)\citenamefont
  {{Lemos}}, \citenamefont {{Lee}}, \citenamefont {{Efstathiou}},\ and\
  \citenamefont {{Gratton}}}]{Lemos19}%
  \BibitemOpen
  \bibfield  {author} {\bibinfo {author} {\bibfnamefont {P.}~\bibnamefont
  {{Lemos}}}, \bibinfo {author} {\bibfnamefont {E.}~\bibnamefont {{Lee}}},
  \bibinfo {author} {\bibfnamefont {G.}~\bibnamefont {{Efstathiou}}}, \ and\
  \bibinfo {author} {\bibfnamefont {S.}~\bibnamefont {{Gratton}}},\ }\href
  {\doibase 10.1093/mnras/sty3082} {\bibfield  {journal} {\bibinfo  {journal}
  {\mnras}\ }\textbf {\bibinfo {volume} {483}},\ \bibinfo {pages} {4803}
  (\bibinfo {year} {2019})},\ \Eprint {http://arxiv.org/abs/1806.06781}
  {arXiv:1806.06781 [astro-ph.CO]} \BibitemShut {NoStop}%
\end{thebibliography}
%

\end{document}